\documentclass[apj]{emulateapj}
\usepackage{color}
\usepackage{natbib}
\usepackage{graphicx}
\usepackage{color}
\usepackage{url}
\usepackage[backref,breaklinks,colorlinks,citecolor=blue]{hyperref}
\usepackage[all]{hypcap}

\newcommand{\msun}{\mbox{$\,{\rm M}_\odot$}}

\shorttitle{Sizes of Globular Clusters as Halo Tracers}

\shortauthors{A. H. Zonoozi,  M. Rabiee, H. Haghi and A. H. W. K\"{u}pper }
\begin{document}

\title{{\bf The Sizes of Globular Clusters as Tracers of Galactic Halo Potentials} }

\author{A. H. Zonoozi \altaffilmark{1},  M. Rabiee \altaffilmark{1} H. Haghi \altaffilmark{1}, A. H. W. K\"{u}pper \altaffilmark{2,3}}
\email{a.hasani@iasbs.ac.ir}
\altaffiltext{1}{Institute for Advanced Studies in Basic Sciences (IASBS), P. O. Box 45195-1159, Zanjan, Iran}
\altaffiltext{2}{Department of Astronomy, Columbia University, 550 West 120th Street, New York, NY 10027, USA}
\altaffiltext{3}{Hubble Fellow}

\begin{abstract}

We present $N$-body simulations of globular clusters, exploring the effect of different galactic potentials on cluster sizes, $r_h$. For various galactocentric distances, $R_G$, we assess how cluster sizes change when we vary the virial mass and concentration of the host galaxy's dark-matter halo. We show that sizes of GCs are determined by the local galactic mass density rather than the virial mass of the host galaxy. We find that clusters evolving in the inner haloes of less concentrated galaxies are significantly more extended than those evolving in more concentrated ones, while the sizes of those orbiting in the outer halo are almost independent of concentration.
Adding a baryonic component to our galaxy models does not change these results much, since its effect is only significant in the very inner halo. Our simulations suggest that there is a relation between $r_h$ and $R_G$, which systematically depends on the physical parameters of the halo. Hence, observing such relations in individual galaxies can put a new observational constraint on dark-matter halo characteristics. However, by varying the halo mass in a wide range of $10^{9}\leq M_{vir}/\msun \leq10^{13}$, we find that the $r_h-R_G$ relationship will be nearly independent of halo mass, if one assumes $M_{vir}$ and $c_{vir}$ as two correlated parameters, as is suggested by cosmological simulations.
\end{abstract}

\section{Introduction}\label{Sec:Intro}

Globular clusters (GCs) are self-gravitating aggregates of tens of thousands to several millions of stars clustered within a few parsecs. As they are among the oldest objects in the Universe, they witnessed most of the history of their host galaxies and may contain valuable fossil records of their host's formation and evolution (e.g., \citealt{Searle78, Kravtsov05, Renaud13}).

The GCs we observe today have survived within, and adapted to, the hostile gravitational potentials of their hosts over billions of years (e.g., \citealt{Fall01, Brockamp14}). With few exceptions, they can be found in basically every galaxy, where their numbers roughly scale with their host's mass and density \citep{Mieske14, Harris15}. Milky-Way-sized galaxies typically show up to a few hundred globular clusters, whereas more massive galaxies can host up to several thousand GCs \citep{Harris13}. They are therefore abundant tracers of galaxy potentials, and have often been used as dynamical probes (e.g, \citealt{Schuberth10, Richtler11, Schuberth12}). The idea of this work is to explore what we can learn from the properties of a galaxy's present-day GC population about the galaxy itself beyond using them as dynamical tracers.

One of the most important and accessible properties of GCs are their sizes. Half-light radii of GCs can be reliably measured out to  several Mpc, giving us a large data set to compare our GC evolutionary models to \citep{Harris13}. About 160 GCs have been identified in the Milky Way (MW), distributed out to 130\,kpc from the Galactic Center \citep{Harris96, Harris10}. Their mean, present-day, half-light radii lie around 3\,pc (e.g., \citealt{vandenBergh12}). In that, they are comparable to the mean sizes for most extra-galactic GCs \citep{Jordan05, Brodie11}.

However, many GCs are significantly more extended than the mean \citep{Bruns12, Norris14}. In the MW, a clear correlation between the sizes of GCs ($r_{hl}$) and their Galactocentric distances ($R_G$) was shown in several observational studies (see \citealt{Hodge60, Hodge62} as pioneering studies). \citet{vandenBergh91} suggested an empirical power-law relation of size versus Galactocentric distance as $r_{hl}\propto\sqrt{R_G}$, independent of GC classification, however, only including GCs out to $R_G\simeq30$ kpc from the Galactic Center.

Such a correlation between $r_{hl}$ and $R_G$ could be primordial, i.e., a result of a correlation between cluster size and galactic tidal field strength and/or gas density at formation (e.g, \citealt{Elmegreen08, Elmegreen10}). Or it could be the result of the preferred disruption of extended GCs near the Galactic Center \citep{Vesperini97, Baumgardt03}. A third alternative would be the expansion of initially compact GCs up to the respective Jacobi radius, which is roughly proportional to $R_G^{2/3}$ for a given GC mass. This is the hypothesis that we are going to test here with the help of $N$-body simulations.

Dynamical evolution plays a key role in shaping characteristics of single globular clusters as well as entire globular cluster systems (e.g., \citealt{Mackey08}). The long-term dynamical evolution of GCs and the consequent mass-loss depend on both physical processes within the cluster as well as on external processes. The most important internal processes are mass-loss from stellar evolution, two-body relaxation and subsequent evaporation, heating from binaries, and ejection of stars caused by few-body interactions (e.g, \citealt{Baumgardt03, Heggie03, Hurley07, Heggie08, Kuepper08, Gieles11}). External influences are mainly coming from changes in the gravitational potential of the host with time, and induced mass-loss through tidal shocks when a cluster traverses the galactic plane, or passes through pericenter on an eccentric orbit \citep{Vesperini97, Gnedin99, Miholics14, Webb14a, Webb14b, Haghi15a}.

As a consequence of dynamical evolution, sizes of GCs can vary significantly over the course of billions of years -- in contradiction with the classical notion that the radius of isolated star clusters remain constant, or change little over a few two-body relaxation times \citep{Spitzer72, Lightman78, Aarseth98}. \citet{Madrid12} showed that a cluster's half-mass radius only remains constant over several relaxation times, when expansion driven by the internal dynamics of the star cluster and the influence of the host galaxy tidal field balance each other. They derived a relation between present-day half-mass radius, $r_h$, and galactocentric distance, $R_G$, of their simulated star clusters, which takes the mathematical form of a hyperbolic tangent, $r_h = r_{h,max}\cdot\tanh(\alpha R_G)$, with a free parameter $\alpha$ determining the slope of the inner part of the relation. The authors showed that the maximum half-mass radius reaches a plateau, $r_{h,max}$, at large galactocentric distances. This is in contrast with the relation found by \citet{vandenBergh91}, which does not include this flattening at large galactocentric distances.

Following the evolution of star clusters at different galactocentric distances, \citet{Haghi14} investigated the impact of primordial mass segregation on the size scale and mass-loss rate of GCs. They showed that initially segregated models undergo a stronger expansion than the unsegregated ones, owing to the rapid mass-loss from the inner part of the star cluster associated with stellar evolution of more massive stars. Furthermore, they found that initially segregated clusters reach significantly larger sizes ($r_{h,max}$) than unsegregated ones (by a factor of 2), suggesting that some of the very extended outer-halo GCs like Palomar\,14 and Palomar\,4 \citep{Jordi09, Frank12, Frank14} may have been born with primordial segregation, and supporting similar conclusions by \citet{Zonoozi11}, \citet{Zonoozi14}, \citet{Haghi15b}, and \citet{Bianchini15}.

The takeaway from \citet{Haghi14} is that the maximum sizes of GCs in a galaxy depend crucially on their birth conditions, which are hard to access -- especially for extra-galactic GCs. Hence, there may be a significant degeneracy between the sizes GCs can reach in a given host gravitational potential and their birth conditions. Breaking these degeneracies will require looking simultaneously at several internal properties of ensembles of GCs, such as masses, sizes, cluster concentrations, mass function slopes, and binary properties. Significant progress has been made by \citet{Leigh13, Leigh15} and \citet{Webb15} in this respect. \citet{Leigh13} showed that the observed correlation between the cluster concentration and present-day mass function slope of Galactic GCs, as reported by \citet{De Marchi07}, cannot be understood by secular long-term dynamical processes alone, and that at least some initial correlation between the initial concentration and the cluster mass is required. \citet{Leigh15} used simulations with the MOCCA Monte Carlo code \citep{Giersz13} to put constraints on the properties of primordial binary populations in Galactic GCs. \citet{Webb15} inferred birth masses of Galactic GCs based on their present-day mass function slopes. Ultimately, such investigations may be able to constrain birth properties of individual GCs well enough such that their present-day sizes can be used as measures for the gravitational potential of their host galaxy.

We will take a different approach here: we ask the question if it is possible to infer information about the gravitational potential of a host galaxy simply by the shape of the size-Galactocentric distance relation of the respective galaxy's GC population (i.e., rather from $\alpha$ than $r_{h,max}$).

In a first study and using $N$-body models of low-mass star clusters ($N=1000$), \cite{Praagman10} found that increasing the mass of a galactic dark-matter halo and its concentration enhances mass-loss rates and, thus, implies shorter dissolution timescales. Based on these first simplistic experiments, we here aim at systematically investigating the effect of the galactic potential, especially the DM halo (described in Sec.~\ref{sec:potential}), on the dynamical evolution of GCs by preparing a comprehensive set of direct $N$-body simulations (Sec.~\ref{sec:models}). We assess how changing the galactic potential influences the mass-loss rate of a star cluster and how this affects the size distribution of star clusters in a galaxy (Sec.~\ref{sec:results}), and draw our conclusions in Sec.~\ref{sec:conclusions}.

\section{The dark matter halo}\label{sec:potential}

In the $\Lambda$CDM picture of structure formation, all galaxies are surrounded by massive haloes of dark matter (DM), which can easily be $>10$ times more massive and extended than the baryonic mass of a galaxy. We therefore focussed on varying the DM halo in our investigation.

In the so-called spherical collapse model \citep{Gunn72}, which is the simplest scenario of the formation of DM haloes, spherically symmetric density perturbations collapse in a homogeneous universe, then virialize and finally reach equilibrium configurations. Inspired by this simple model, it is conventional to describe a DM halo as a spherical region with density of about 200 times the critical density of the universe, $\rho_{crit} = 3 H^2(z)/8 \pi G$. Hence the virial mass of this spherical over-dense region is given by:
\begin{equation}
M_{vir} = \frac {4}{3} \pi \Delta_{vir} \rho_{crit} r_{vir}^3 ,
\end{equation}
where $\Delta_{vir}$ is the virial overdensity criterion, which is a function of cosmology and redshift and
varies from 100 to 200. $r_{vir}$ is the virial radius of a sphere that encloses an average
density of $\Delta_{vir}$ $\times$ $\rho_{crit}$  within the virialized region.

Cosmological simulations of collisionless DM particles suggest that equilibrium DM haloes, produced through hierarchical clustering, are well approximated by a universal, two-parameter density profile (\citealt{Navarro97}; hereafter NFW) as
\begin{equation}\label{eq:nfwprofile}
 \rho_{NFW}(r) = \frac {\rho_0}{r/r_s (1+r/r_s)^2},
\end{equation}
where $\rho_0$ and $r_s$ are scaling parameters that characterize a given halo. By integrating Eq.~\ref{eq:nfwprofile}, the total mass inside the radius $r$ is given by
\begin{equation}
 M_{NFW}(r) = 4 \pi \rho_0 r_s ^3 \left[\ln(1+r/r_s)-\frac{r/r_s}{1+r/r_s}\right].
\end{equation}
The NFW profile can be equivalently identified by the virial mass, $M_{vir}$, and concentration, $c = r_{vir}/r_s$, which relates the inner and virial parameters as
\begin{eqnarray}
 \rho_0 &=& \frac{\rho_{crit} \Delta_{vir}}{3} \frac{c^3}{\ln(1+c) - c/(1+c)},\\
 r_s &=& \frac{1}{c} \left(\frac{3 M_{vir}}{4\pi \Delta_{vir} \rho_{crit}}  \right)^{1/3}.
\end{eqnarray}
The potential to which the NFW density profile corresponds is given by
\begin{equation}
\Phi(r) = - \frac {G M_{vir}}{r}  \frac{\ln(1+r/r_s)}{\ln(1+c)- c/(1+c)}.
\end{equation}
Large simulations of structure formation have shown that there exists a (redshift-dependent) correlation between halo concentration, $c$, and the virial mass, $M_{vir}$, such that more massive halos are less concentrated \citep{Bullock01, Wechsler02, Neto07, Klypin11}. The concentration-mass relation for relaxed halos can be approximated as
\begin{equation}\label{eq:c-mvir-relation}
\log_{10}\left( c\right) = 1.025-0.097\,\log_{10}\left(\frac{M_{vir}}{10^{12}h^{-1}\msun}\right)\label{c-mvir}.
\end{equation}
Therefore, the NFW halo profile can be rewritten in terms of a single parameter, $M_{vir}$. In Sec.~\ref{ssec:oneparameter} we will show what such a correlation between virial mass and concentration implies for the scale sizes of globular clusters.

\begin{table*}
\centering
\caption{Properties of the NFW halo models and summary of the initial and final parameters of the $N$-body runs presented in Sec.~\ref{ssec:NFW}. The initial mass and half-mass radius of all models are $M=10^5\msun$ and $r_h=6$\,pc, respectively. Column~1 gives the galactocentric distance of each model. $M_{tot}(R_G)$  is the total enclosed mass of the galaxy within $R_G$. Column~3 gives the mean density of the galaxy within $R_G$, defined as $3M_{tot}(R_G)/(4\pi R_G^3)$. Column~4  is the initial filling factor, where $r_{h0}= 6$\,pc for all models. Columns 5--8 contain the parameters of the NFW halo model.  The following columns give the 3D half-mass radius and final mass of the simulated star clusters after 13\,Gyr of evolution.}
\begin{tabular}{ccccccccccccc}
\hline
(1)&(2)&(3)&(4)&(5)&(6)&(7)&(8)&(9)&(10)\\
\hline
$R_{G}$ & $M_{tot}$          & $\bar{\rho}_G$          &  $f_0$ & $M_{vir}$   & $c$ & $r_{vir}$  & $r_{s}$ & $r_{h,f}$ &  $M^{f}$ \\
\,[kpc] &[$10^{10}$M$_\odot$]  & [M$_\odot pc ^{-3}$]& $[r_h/r_t]_0$ & [M$_\odot$] &           & [kpc]      & [kpc]   & [pc]      & [$10^3$M$_\odot$]\\
\hline
2     &0.1  & 3.6$\times 10^{-2}$     & 0.084     &$10^{12}$ & 5 & 200 & 40 &14.9& 16 \\
8.5   &1.8  &7.0$\times 10^{-3}$      & 0.051     &$10^{12}$ & 5 & 200 & 40 & 18.0 & 21 \\
50    &26.0 & 5.1$\times 10^{-4}$    & 0.024      &$10^{12}$ & 5 & 200 & 40 &20.2& 24\\
\hline
2     &0.3  & 8.8$\times 10^{-2}$    & 0.115     &$10^{12}$ & 10 & 200 & 21 &9.5& 5   \\
8.5   &3.7  & 1.4$\times 10^{-2}$   & 0.068     & $10^{12}$ & 10 & 200 & 21 & 16.3 & 18\\
50    &36.0 & 6.9$\times 10^{-4}$    & 0.028     &$10^{12}$ & 10 & 200 & 21 &20.1& 24\\
\hline
2     & 0.7 & 2.2$\times 10^{-1}$   & 0.163     &$10^{12}$ & 20  & 200 & 10  &--& -- \\
5     & 3.4 & 6.5$\times 10^{-2}$    & 0.115     &$10^{12}$ & 20  & 200 & 10  & -- & -- \\
8.5   & 7.4 & 2.9$\times 10^{-2}$    & 0.091     &$10^{12}$ & 20  & 200 & 10  &12.3& 10 \\
15    &15.0 & 1.1$\times 10^{-2}$    & 0.068     &$10^{12}$ & 20  & 200 & 10  &16.4&  16 \\
50    &45.6 &8.7$\times 10^{-4}$     & 0.031     &$10^{12}$ & 20  & 200 & 10  &19.4 & 23\\
100   &71.0 & 1.7$\times 10^{-4}$    & 0.018     &$10^{12}$ & 20  & 200 & 10  & 19.5 & 25\\
\hline
\end{tabular}
\label{tab-nfw}
\end{table*}

\begin{figure*}
\begin{center}
\includegraphics[width=58mm]{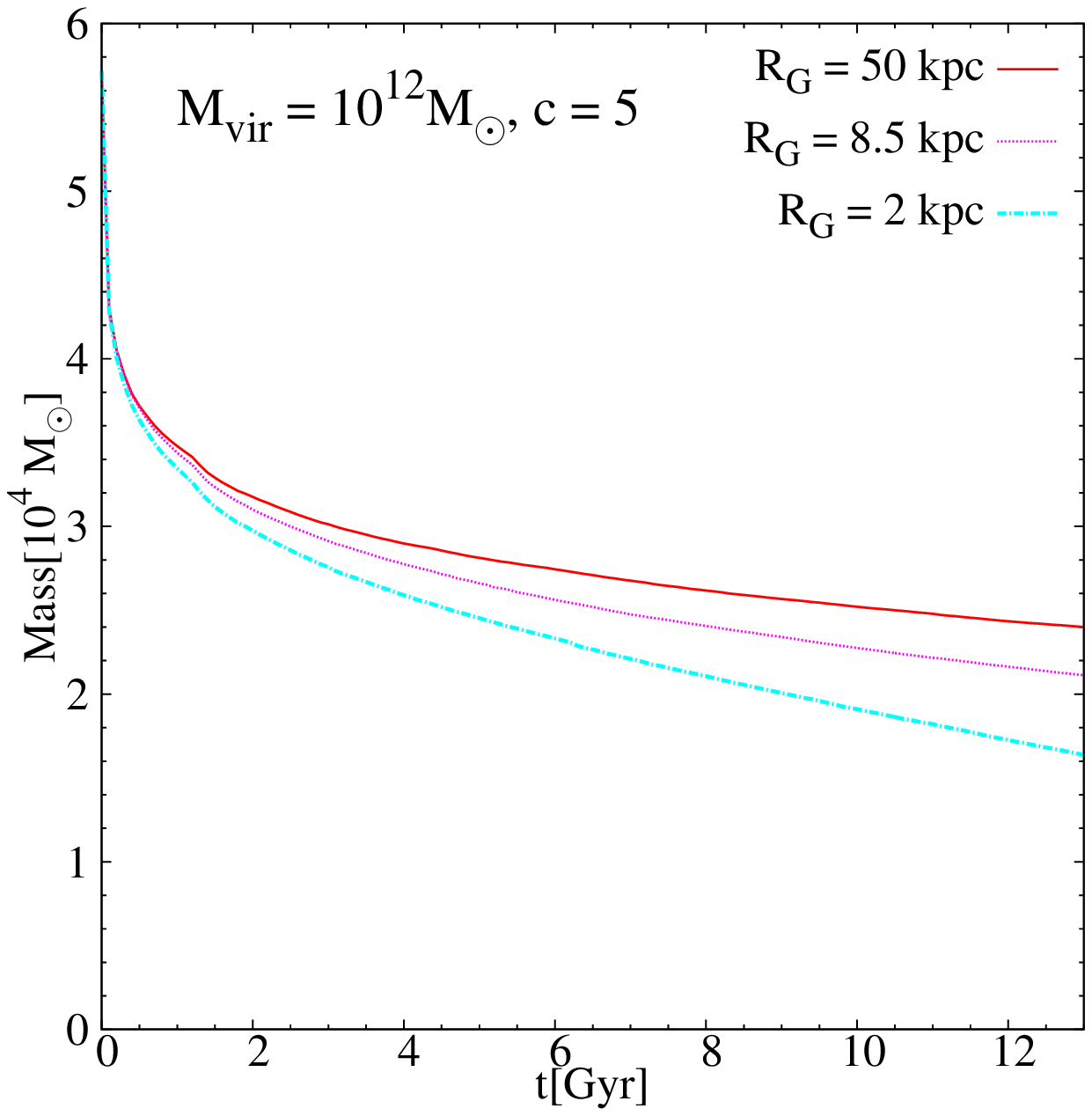}
\includegraphics[width=58mm]{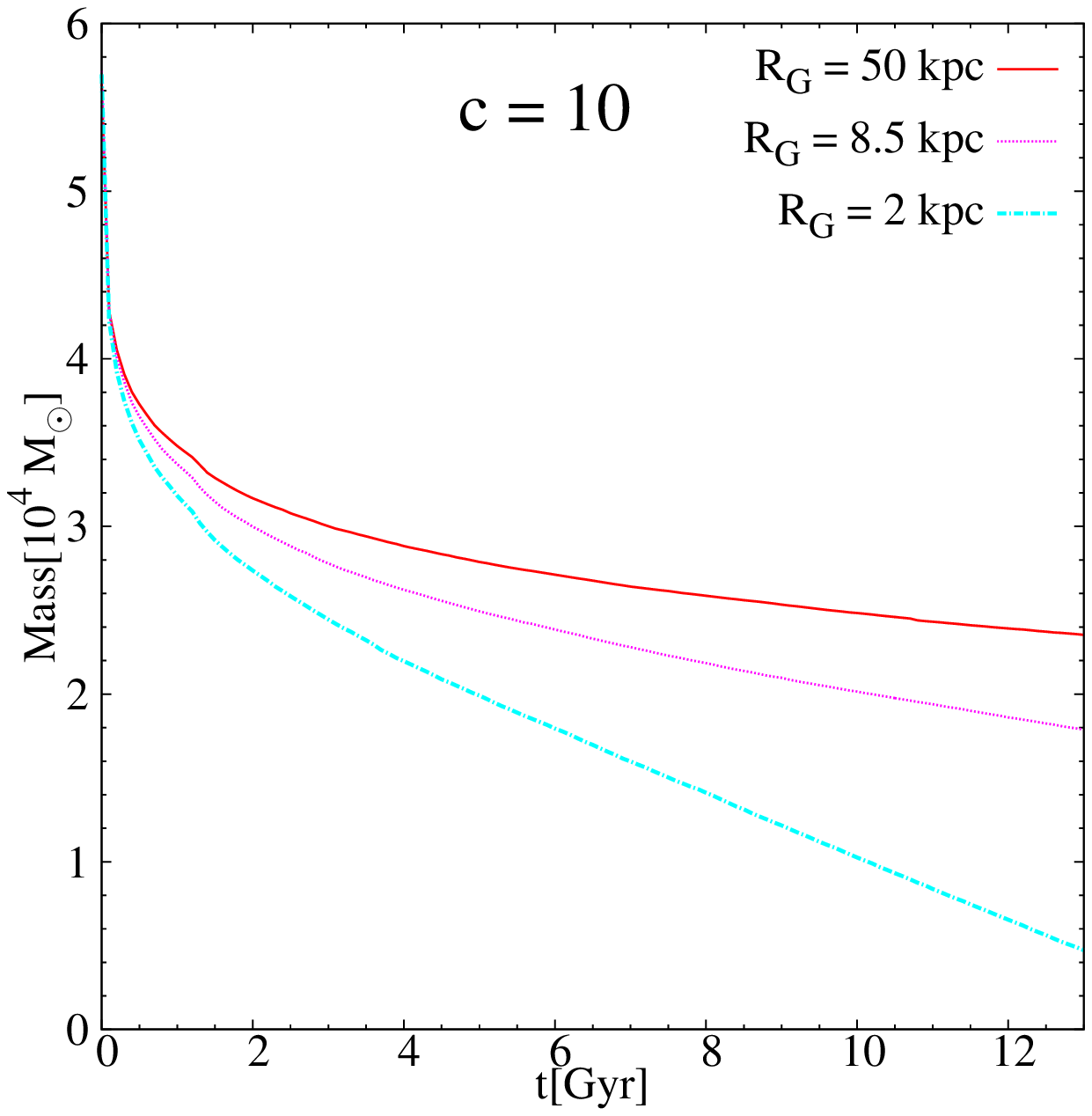}
\includegraphics[width=58mm]{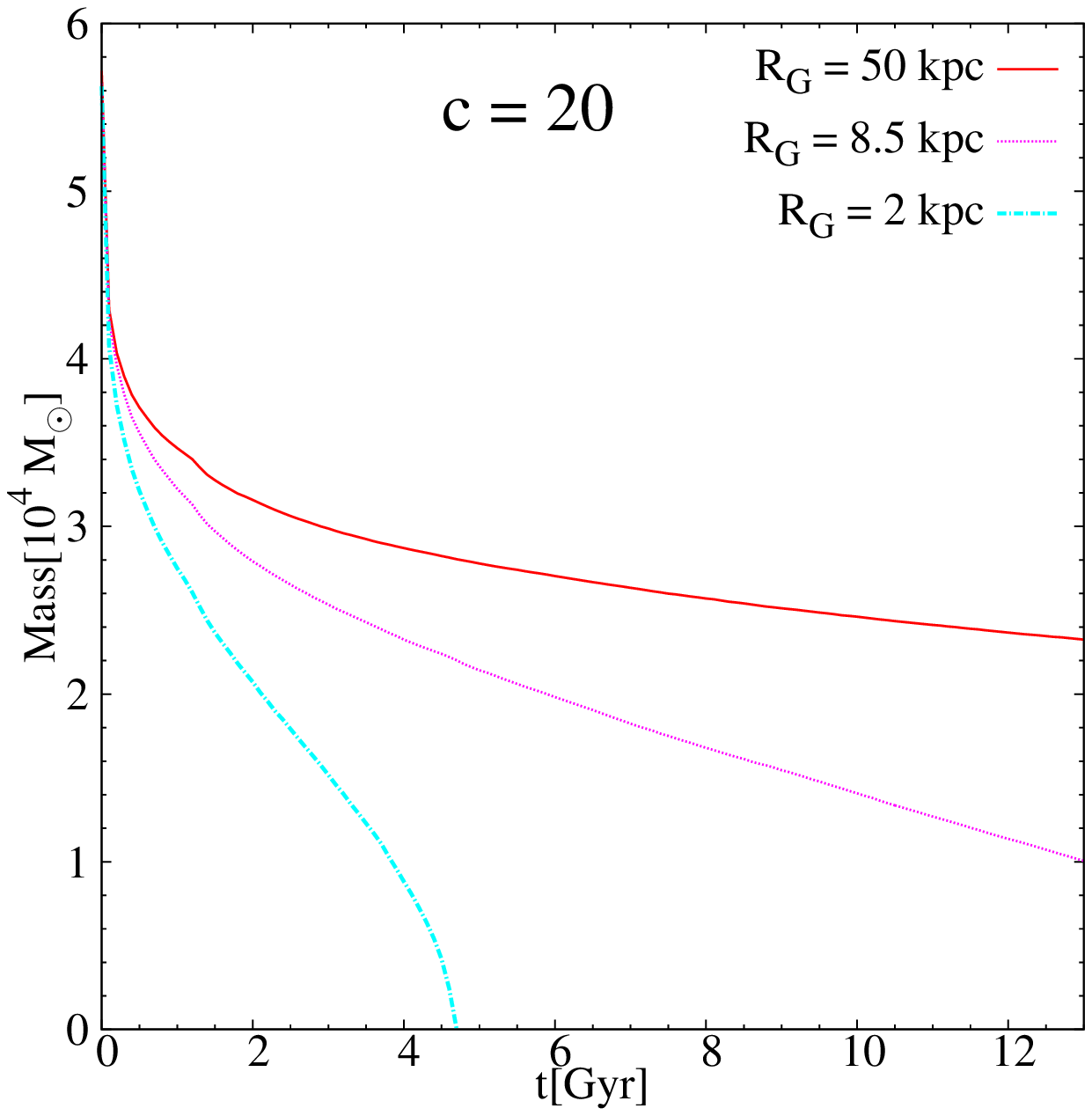}\\
\includegraphics[width=58mm]{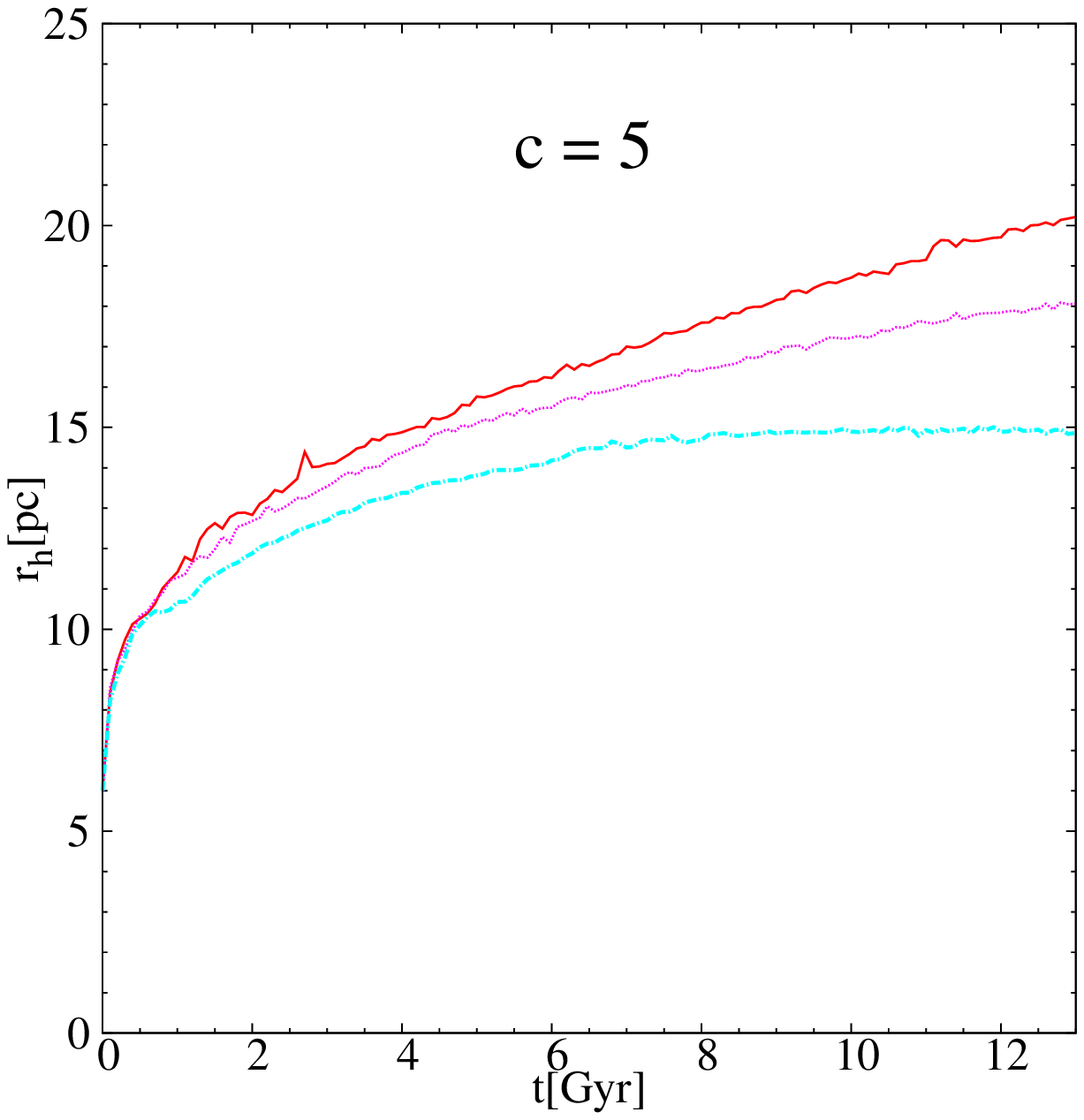}
\includegraphics[width=58mm]{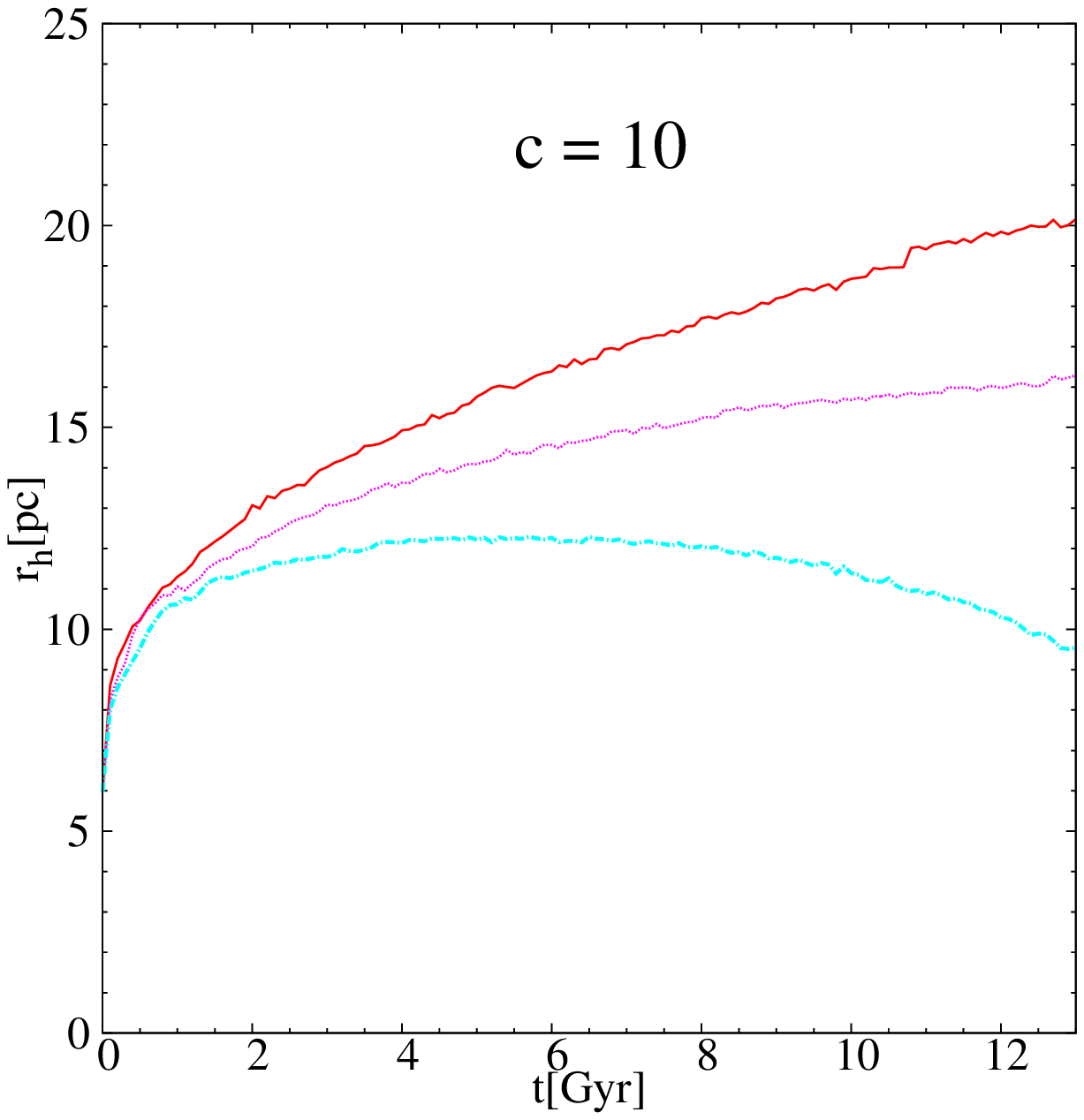}
\includegraphics[width=58mm]{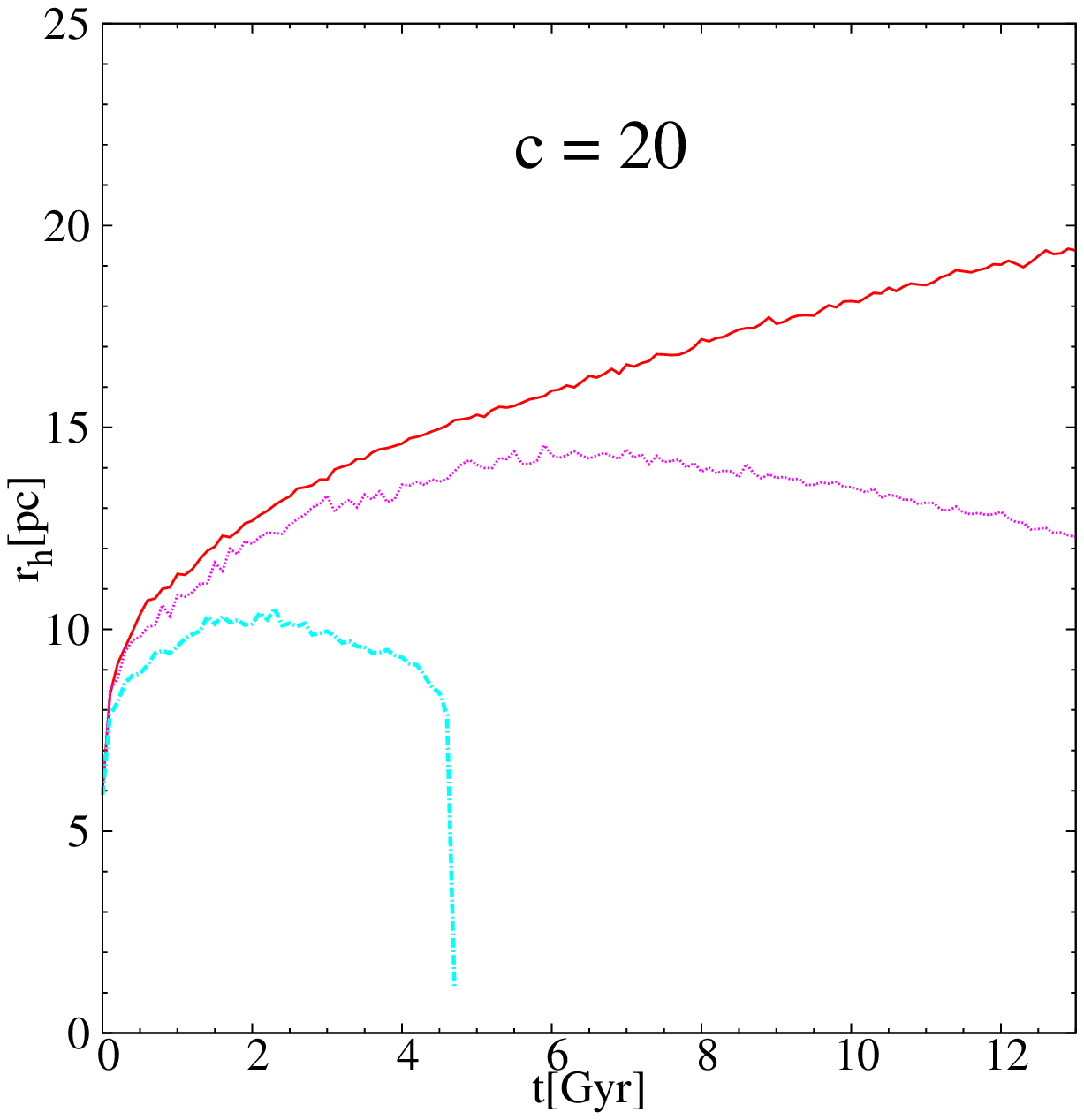}
\caption{Evolution of the total mass (top panels) and 3D half-mass radius (bottom panels) of the clusters at different orbital distances in the range $R_G=2-50$\,kpc from the galaxy center for 3 different halo concentration parameters, $c$. The halo viral mass is the same for all models ($M_{vir}=10^{12}\msun$). All clusters are starting with an initial 3D half-mass radius of 6\,pc and $N=10^5$ stars. The more concentrated halo enhances the mass loss rate and limits the expansion of orbiting star clusters owing to a larger mean enclosed mass density within the orbital radius, resulting in a stronger tidal field. This effect is significantly more evident for the clusters close to the galactic center. Enhanced mass loss rates imply shorter dissolution times and smaller final sizes for the surviving clusters.}
\label{ml-NFW}
\end{center}
\end{figure*}

\begin{figure}
\centering
\includegraphics[width=85mm]{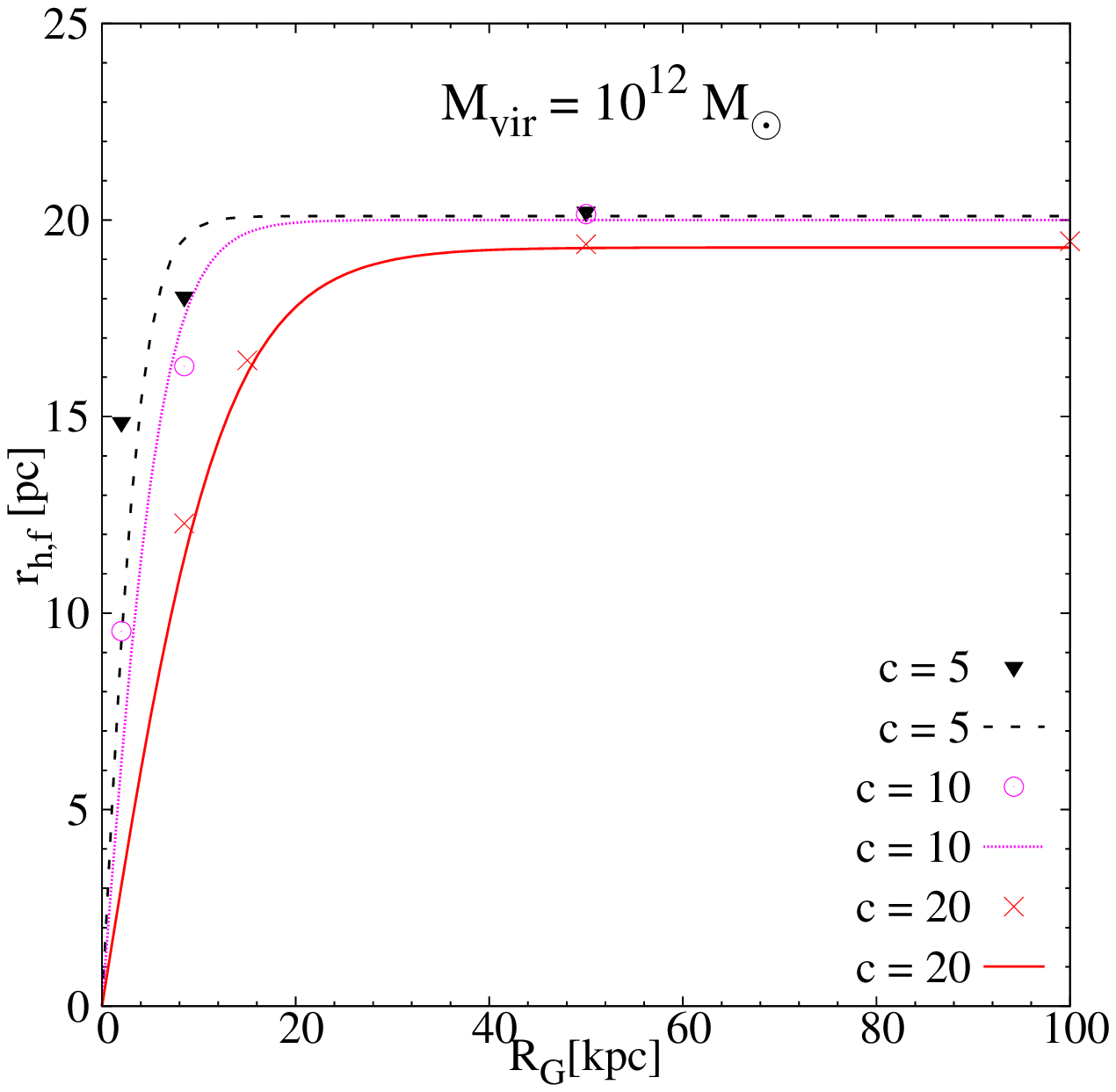}
\caption{Final three-dimensional half-mass radii after a Hubble time of evolution of the simulated star
clusters versus galactocentric distances. For all halo models we use $M_{vir}=10^{12}\msun$ but vary the halo concentrations (Sec.~\ref{ssec:NFW}). The symbols depict the values for three different halo concentrations. The lines are the best-fitting functions in the form of a hyperbolic tangent as given by Eq.~\ref{eq:rh-RG}. The red solid line, pink dotted line, and black dashed line correspond to models with halo concentrations of 20, 10, and 5, respectively.  The maximum size, $r_{h,max}$, of the modeled clusters are nearly the same for all galaxy models, but the inner slope (within the inner 20\,kpc) is different for different concentrations. The best-fitting parameters we obtain for each halo model are given in Table \ref{fit-nfw}.}
\label{Rh-Rg}
\end{figure}

\section{$N$-body models}\label{sec:models}

Our simulations to study the evolution of star clusters in different galactic potentials were carried out on desktop workstations with Nvidia 690 Graphics Processing Units at the Institute for Advanced Studies in Basic Sciences (IASBS), using the state-of-the-art collisional $N$-body integration code \textsc{Nbody6} \citep{Aarseth03, Nitadori12}. To follow the orbits of stars, it uses a fourth-order Hermite scheme with individual time-steps. The code also invokes regularization schemes to deal with the internal evolution of binaries and multiples, and also to account for close encounters.  Stellar and binary evolution are included in the code by using the \textsc{SSE/BSE} packages based on analytical fitting functions  developed by \citet{Hurley00}, \citet{Hurley02}, and \citet{Hurley05}.

We used the publicly available code \textsc{McLuster}\footnote{\url{https://github.com/ahwkuepper/mcluster.git}} to generate our initial conditions for the star clusters \citep{Kuepper11}. All models in this study start with $N_i=10^5$ particles and a total initial mass of $5.7 \times 10^4 \msun$. We use a broken power-law initial mass function (IMF, \citealt{Kroupa01}) with a slope of $\alpha = 1.3$ for stars lighter than $0.5\msun$ and $\alpha = 2.3$ for stars above this mass in the range between the limits of $m_{min} = 0.08 \msun$ and $m_{max} = 100\msun$. The modeled clusters have a metallicity of $Z = 0.001$ or [Fe/H]\,$\approx-1.3$.

Stars are initially distributed following a Plummer radial density profile \citep{Plummer11} with an initial half-mass radius of 6\,pc for all modeled clusters. Their initial velocities are generated based on the assumption of virial equilibrium.
We assume that, at the beginning of the simulations, the residual gas, left over from the star formation process, is completely expelled from the clusters and stars are on the zero age main sequence. Stars beyond twice the tidal radius are counted as escaped stars and removed from the computations.

In this work, all models are orbiting on circular orbits and have exactly the same initial set up with the exception of the initial galactocentric distance and the properties of the galactic potential. We created three sets of models:
\begin{enumerate}
\item In the first set of models, we study the evolution of star clusters in three different static, spherical NFW haloes with the same virial mass, $M_{vir}=10^{12}\msun$, resembling approximately the mass of our own Galaxy, but different concentrations, $c=5, 10, 20$. This first test case has no baryonic component like a disk or bulge. For each adopted concentration, identical star clusters are placed on circular orbits at $R_G = 2, 8.5$ and 50\,kpc from the galactic center, representing the inner, intermediate, and outer part of the galaxy. From this first set we gain an intuition on how the evolution of the clusters change when we change the shape of the DM halo.
\item In the second set, the clusters orbit in a more Milky Way-like potential consisting of three components: bulge, disk and DM halo. The bulge is modeled as a central point-mass:
\begin{equation}
 \Phi_b \left(r\right) = -\frac{G M_b}{r},
\end{equation}
where $M_b$ is the mass of the bulge component. The disk component of the galaxy is modeled following the prescriptions of \citet{Miyamoto75}:
\begin{equation}
\Phi_d \left(x,y,z\right) = -\frac{G M_d}{\sqrt{x^2+y^2+\left(a + \sqrt{z^2+b^2}\right)^2}},
\end{equation}
where $a$ is the disk scale length, $b$ is the disk scale height, and $M_d$ is the total mass of the disk component. We used values of $a= 4$\,kpc and $b=0.5$\,kpc, while for the disk and bulge masses we adopted $M_d = 5\times 10^{10}\msun$ and $M_b = 1.5 \times 10^{10}\msun$, respectively, as suggested by \citet{Xue08} for a Milky Way-like potential. We use the same DM halo as in the first setup ($M_{vir}=10^{12}\msun$) and additionally compare this configuration to a more massive halo of $M_{vir}=10^{13}\msun$. As above, we use a model grid of three concentrations for the lower-mass halo, and two different concentration (c=7.5, 25) for the more massive halo. In both galactic potentials we run simulations at six galactocentric distances, $R_G = 5, 10, 15, 30, 50$ and 100\,kpc. This set of models will tell us how for a given baryonic configuration of a galaxy, the DM halo properties influence the evolution of the GCs.
\item In the third set of models, we vary the DM halo configuration using the concentration-virial mass relation from cosmological simulations (Eq.~\ref{eq:c-mvir-relation}). We vary the NFW profile following this relation with different halo masses, i.e. $M_{vir}= 10^9\msun, 10^{11}\msun, 10^{12}\msun$, and $10^{13}\msun$, and leave the baryonic component unchanged. This last set is going to show us how strong the degeneracy between $M_{vir}$ and $c$ is.
\end{enumerate}
Summaries of all models can be found in Tab.~\ref{tab-nfw}, \ref{tab-2} and \ref{tab-1}.

\section{Results}\label{sec:results}

\begin{table*}
\centering
\caption{The initial and final parameters of $N$-body runs and parameters of the dark matter halo as in Table~\ref{tab-nfw}, but for the two-component NFW models (Sec.~\ref{ssec:NFW-2}). }
\begin{tabular}{ccccccccccccc}
\hline
(1)&(2)&(3)&(4)&(5)&(6)&(7)&(8)&(9)&(10)\\
\hline
$R_{G}$ & $M_{tot}$            &  $\bar{\rho}_G$    &  $f_0$        & $M_{vir}$   & $c$ & $r_{vir}$  & $r_{s}$ & $r_{h,f}$ &  $M_{f}$ \\
\,[kpc] & [$10^{10}$M$_\odot$] & [M$_\odot pc ^{-3}$] & $[r_h/r_t]_0$ & [M$_\odot$] &            & [kpc]     & [kpc]   & [pc]      & [$10^3$M$_\odot$]\\
\hline
5     & 4.2   & 8.1$\times 10^{-2}$     & 0.138     &$10^{12}$ & 5 & 200 & 40 &--& --  \\
10    &7.7    & 1.8 $\times 10^{-2}$  & 0.086      &$10^{12}$ & 5 & 200 & 40   &12.15 & 1 \\
15    & 10.6  & 7.5 $\times 10^{-3}$  & 0.063     &$10^{12}$ & 5 & 200 & 40    &15.96& 16 \\
30    & 19.9  & 1.8 $\times 10^{-3}$  & 0.038    &$10^{12}$ & 5 & 200 & 40  &18.31& 23  \\
50    & 32.9  & 6.3 $\times 10^{-4}$  &0.027   &$10^{12}$ & 5 & 200 & 40    &19.71 & 24\\
100   & 62.4  & 1.5 $\times 10^{-4}$  & 0.017 &$10^{12}$ & 5 & 200 & 40    &20.01 & 25 \\
\hline
5     & 5.1    & 9.7$\times 10^{-2}$      & 0.143     &$10^{12}$ & 10 & 200 & 20 &--& --\\
10    & 10.1   & 2.4$\times 10^{-2}$     & 0.092    &$10^{12}$ & 10 & 200 & 20 &11.28& 7  \\
15    & 14.6   & 1.0$\times 10^{-2}$      & 0.069    &$10^{12}$ & 10 &200 & 20  &14.53 &  14 \\
30    & 27.4   & 2.4$\times 10^{-3}$      & 0.043   &$10^{12}$ & 10 & 200 & 20   &18.07& 21 \\
50    & 42.4   & 8.1$\times 10^{-4}$        &0.030    &$10^{12}$ & 10 & 200 & 20  &20.59& 22\\
100   & 70.6   & 1.7$\times 10^{-4}$        &0.018&  $10^{12}$ & 10 & 200 & 20 & 20.09 & 24 \\
\hline
5     & 6.3    & 1.3$\times 10^{-1}$      & 0.157     &$10^{12}$ & 20 & 200 & 10  &--& --\\
10    & 14.4   & 3.4$\times 10^{-2}$      & 0.104     &$10^{12}$ & 20 & 200 & 10   &8.08& 3 \\
15    & 20.9   & 1.5$\times 10^{-2}$      & 0.079     &$10^{12}$ & 20 & 200 & 10   &13.61&  13 \\
30    & 36.6   & 3.2$\times 10^{-3}$     & 0.048  &$10^{12}$ & 20 & 200 & 10  & 17.68 & 21\\
50    & 52.1   & 9.9$\times 10^{-4}$      & 0.033  &$10^{12}$ & 20 & 200 & 10  &19.73 & 23\\
100   & 77.5   &1.8$\times 10^{-4}$      & 0.019 &$10^{12}$ & 20 & 200 & 10  &  20.15 & 25 \\
\hline
5     & 6.2     & 1.2$\times 10^{-1}$      & 0.149     &$10^{13}$ &7.5 & 432 & 58  &--& -- \\
10    & 14.9    & 3.6$\times 10^{-2}$      & 0.098    &$10^{13}$ &7.5 & 432 & 58  &10.31& 6 \\
15    & 25.6    & 1.8$\times 10^{-2}$      & 0.077     &$10^{13}$ &7.5 & 432 & 58   &14.85&  14 \\
30    & 67.3    & 5.9$\times 10^{-3}$      & 0.053    &$10^{13}$ &7.5 & 432 & 58   &17.87& 20\\
50    &  133.0  &2.5$\times 10^{-3}$     & 0.040   &$10^{13}$ &7.5 & 432 & 58   &18.64& 21\\
100   &  301.0  &7.2$\times 10^{-4}$    & 0.028   &$10^{13}$ &7.5 & 432 & 58   &19.64& 23\\
\hline
5     & 16.4    & 3.1$\times 10^{-1}$     & 0.223    &$10^{13}$ & 25 & 432 & 17  &--& -- \\
10    & 44.4    & 1.1$\times 10^{-1}$   & 0.140     &$10^{13}$&25 & 432 & 17  &--& --\\
15    & 75.5    & 5.3$\times 10^{-2}$   & 0.113      &$10^{13}$ &25 & 432 & 17  &--& --\\
20    & 107.0   & 3.2$\times 10^{-2}$   & 0.097   &$10^{13}$ &25& 432 & 17  &10.60 & 7\\
30    & 168.0   & 1.5$\times 10^{-2}$   & 0.077   &$10^{13}$ &25& 432 & 17  &14.05&14\\
50    & 274.0   &5.2$\times 10^{-3}$ & 0.056  &$10^{13}$ & 25 & 432 & 17  &16.87 & 19\\
100   & 468.0   &1.1$\times 10^{-3}$  & 0.034  &$10^{13}$ & 25 & 432 & 17  &19.64 & 23\\
\hline
\end{tabular}
\label{tab-2}
\end{table*}

\begin{figure*}
\centering
\includegraphics[width=85mm]{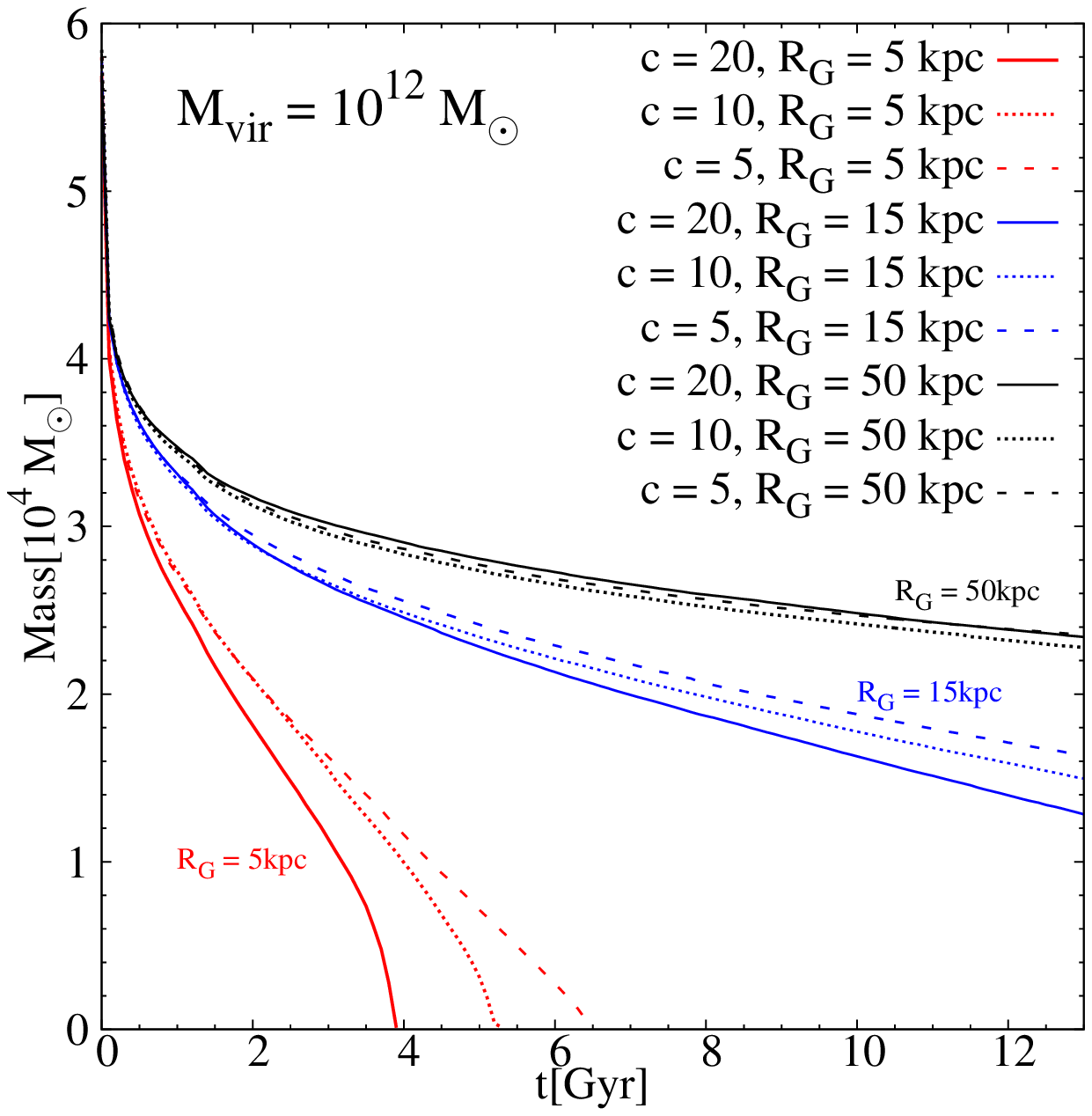}
\includegraphics[width=85mm]{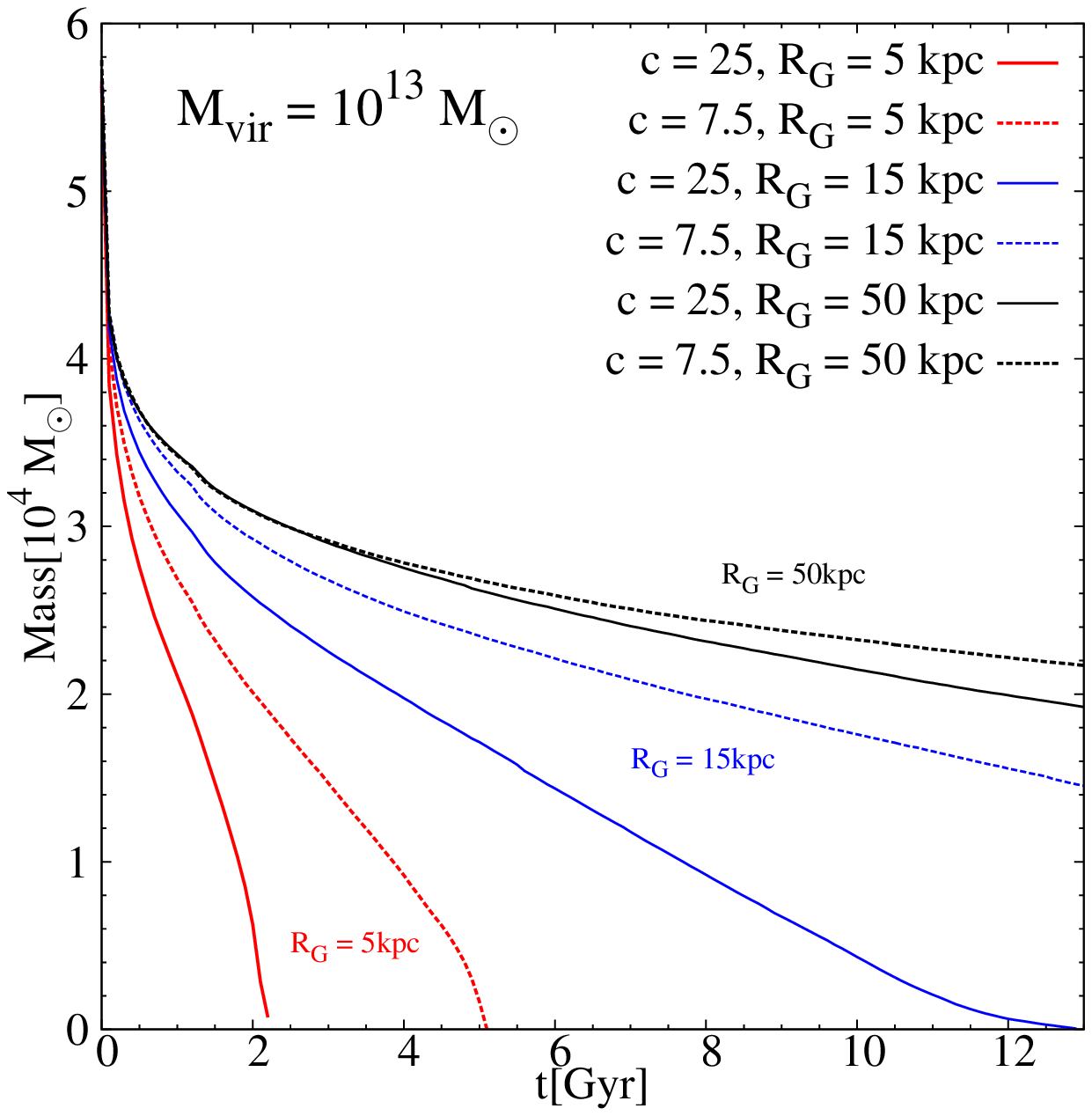}\\
\includegraphics[width=85mm]{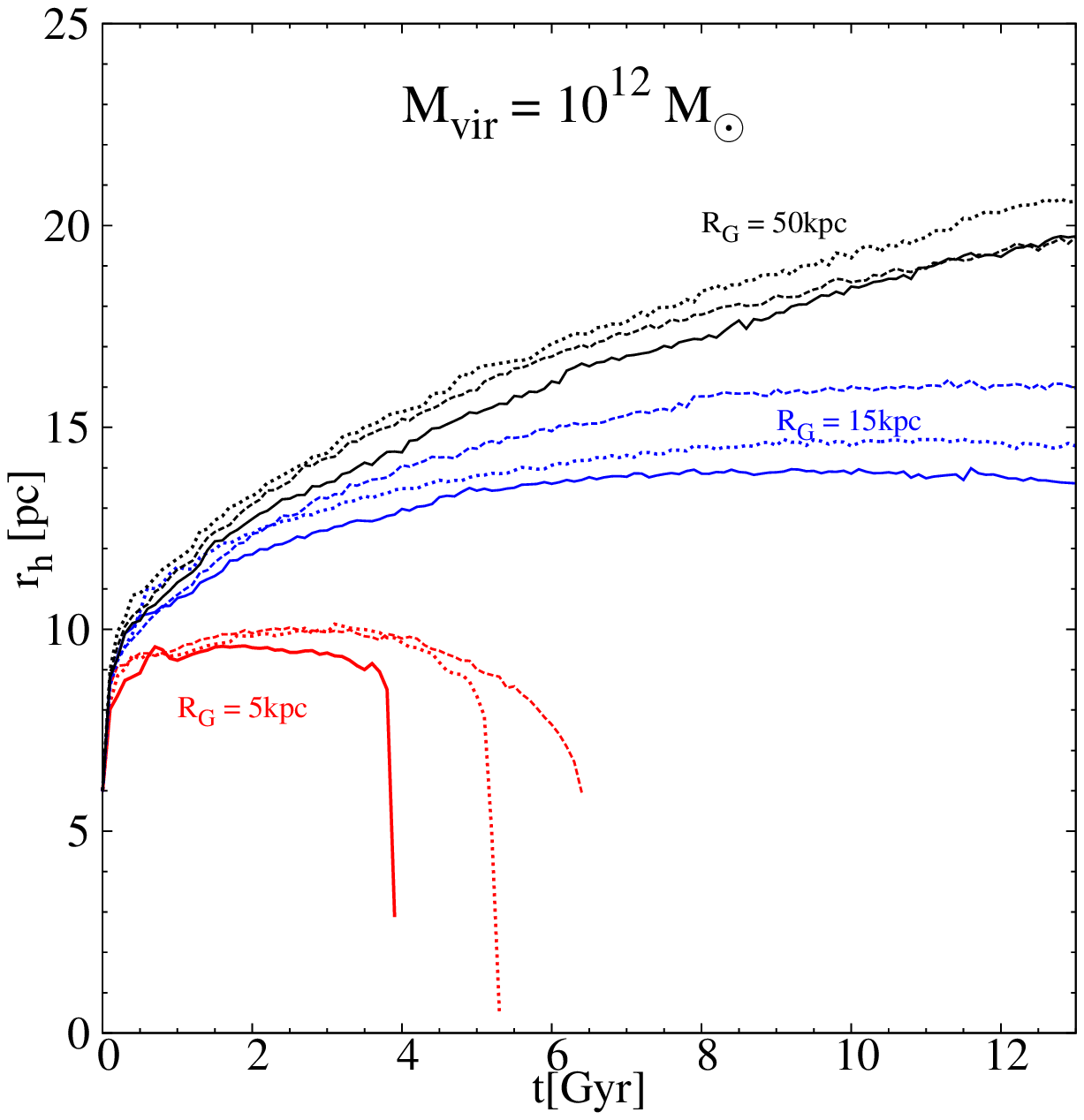}
\includegraphics[width=85mm]{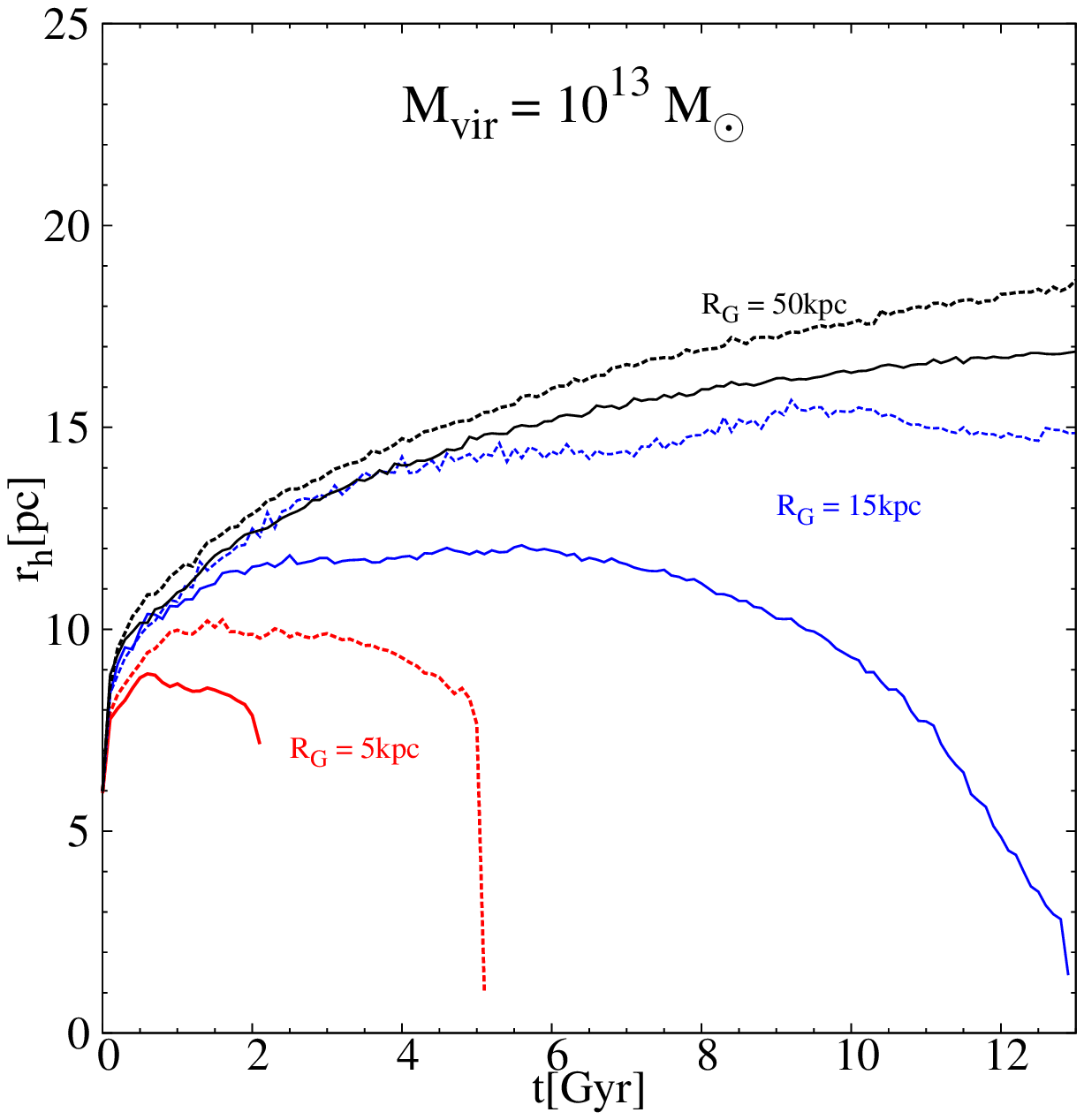}
\caption{Mass-loss and 3D half-mass radius evolution with time as in Fig.~\ref{ml-NFW}, but here all models are evolved within a three-component galaxy model including bulge, disk and halo. The virial mass was chosen as $M_{vir}=10^{12}\msun$ (left panels) and $M_{vir}=10^{13}\msun$ (right panels).  All clusters have an initial 3D half-mass radius of 6 pc.  Only a subset of the simulations is plotted to preserve the clarity of the figure. Top panels: the evolution of the total mass for different galactocentric distances, $R_G$.  Bottom panels: evolution of the 3D half-mass radius of the same models. Different concentrations are chosen for the galaxy haloes. As expected, the presence of a stellar component plays only a dominant role for clusters evolving in the inner part of the galaxies (cf., Fig.~\ref{ml-NFW}).
} \label{ml-Rh-2}
\end{figure*}

\begin{figure*}
\centering
\includegraphics[width=85mm]{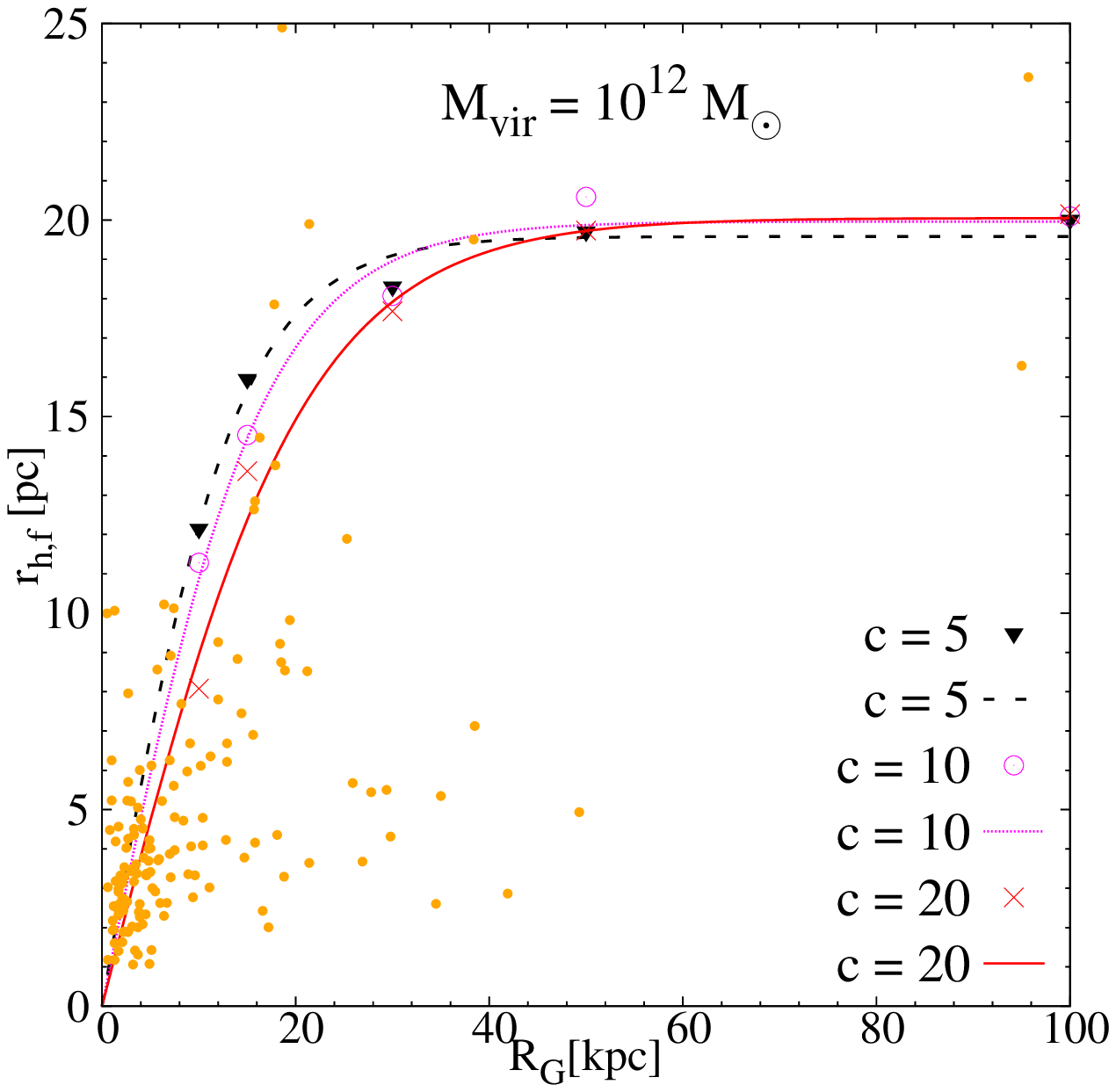}
\includegraphics[width=85mm]{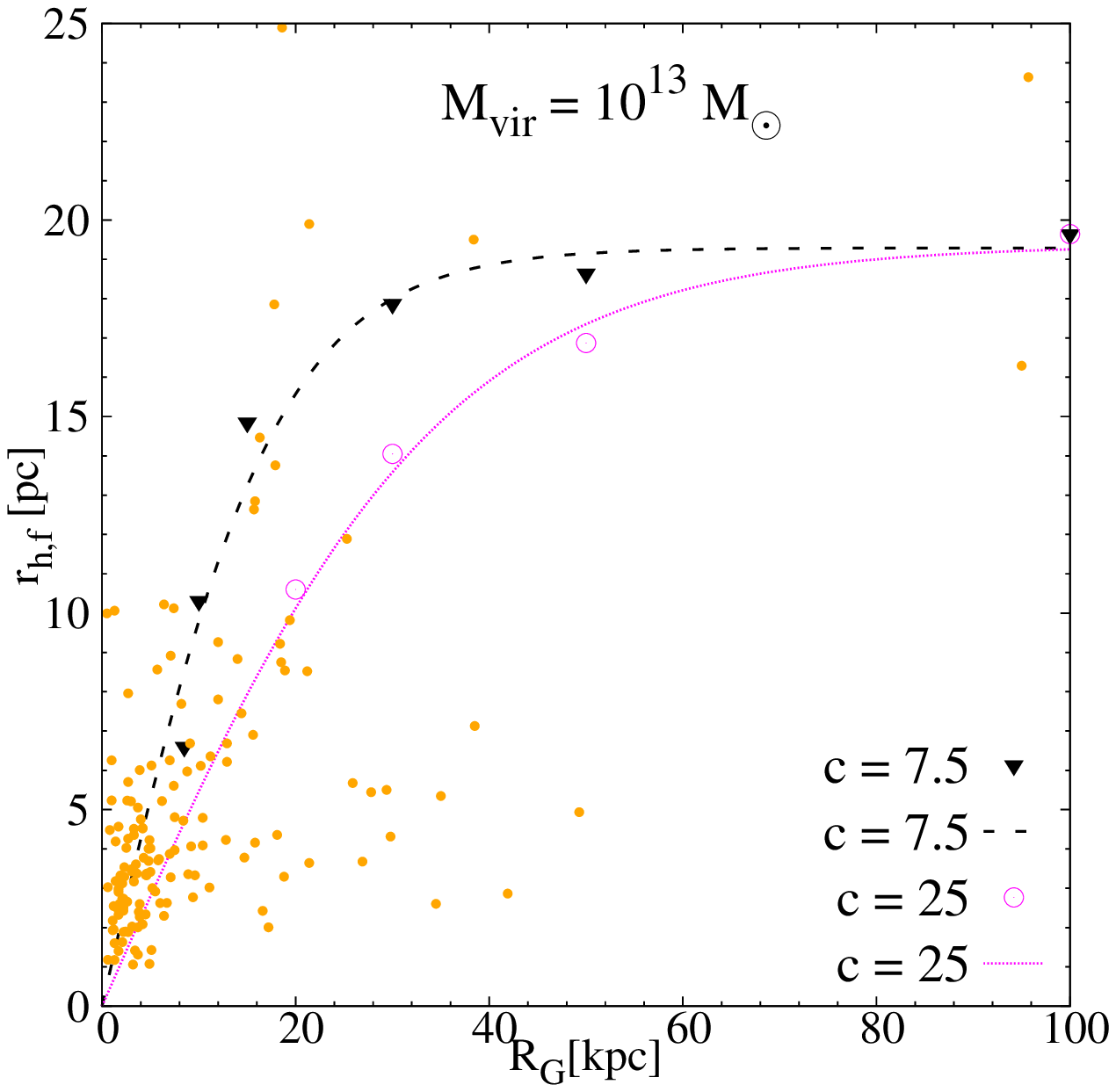}
\caption{Final three-dimensional half-mass radii of the simulated star
clusters after a Hubble time of evolution versus their galactocentric distances. Left: a galaxy with $M_{vir} = 10^{12}\msun$. Right: a galaxy with $M_{vir} = 10^{13}\msun$. The lines are the best-fitting functions given by equation \ref{eq:rh-RG}. The best-fitting parameters that we obtained for each halo model are given in Table \ref{fit-nfw}. 3D half-mass radii (deprojected from 2D effective radii) and Galactocentric distances of MW GCs taken from the Harris catalogue (2010) are over-plotted as filled circles for comparison. GCs on eccentric orbits and tidally underfilling clusters complicate the interpretation of the observational data.}
\label{Rh-Rg-2}
\end{figure*}

In the discussion of the models we will focus on the evolution of the bound mass and the half-mass radius of all bound stars, since those are the two most accessible empirical quantities of GCs, especially for extra-galactic GCs.
We present our three sets of models in the following order: first we show the set in which the galactic potential consists of only an NFW halo (without a baryonic component; Sec.~\ref{ssec:NFW}), then the set within a three-component galactic potential, consisting of a bulge, a disk and an NFW halo (Sec.~\ref{ssec:NFW-2}). Finally, we present the set of simulations of clusters moving in a 3-component galaxy model including a one-parameter NFW halo using the $M_{vir}-c$ relation (Eq. \ref{eq:c-mvir-relation}; Sec.~\ref{ssec:oneparameter}).

\subsection{Evolution in NFW halos with different concentrations} \label{ssec:NFW}

In order to assess the pure effect of the galactic dark matter halo potential on the evolution of GCs, we first calculate clusters orbiting in an NFW halo alone without any baryonic component (model set 1). For the case of $c=20$ we calculate the evolution of our star cluster at three additional radii from the galactic center ($R_G=$ 5, 15, and 100 kpc) to show the dependence of the final half-mass radius on galactocentric distance in more detail.

Figure~\ref{ml-NFW} shows the evolution of the mass and the half-mass radius\footnote{We derived the value of the 3D half-mass radius by taking all stars within the Jacobi radius, and searching for the radius containing half of the total mass.} of our modeled clusters. The general mass loss is due to stellar evolution of massive stars, and evaporation due to two-body relaxation. The mass loss is initially the same for all models since in the first $\approx500$\,Myr of evolution, the mass-loss is dominated by stellar evolution of the most massive stars, which goes along with a rapid expansion of the half-mass radius. Afterwards, tidal mass loss driven by two-body relaxation becomes the dominant channel of cluster mass-loss. In the inner 10\,kpc of the host galaxy, where the clusters are tidally filling this mass-loss leads to the truncation on the clusters' sizes, while in the outer part of the galaxy the clusters keep expanding.

Figure~\ref{ml-NFW} shows that the simulated star cluster evolving at $R_G=2$\,kpc from the center of a galaxy with $c=20$ is completely dissolved at $t\simeq 4.6$\,Gyr, while the same star cluster evolving in a less concentrated galaxy with $c=5$, has a remaining mass of $16000\msun$ at $t=13$\,Gyr ($\simeq 25\%$ of its initial mass). As we will show in Sec.~\ref{sec:discussion}, this is due to the mean mass density of the dark matter halo within the cluster's orbit being larger in the more concentrated halo, resulting in a greater tidal force on the cluster. The influence of concentration is negligible for the clusters evolving in the outer part of the galaxy ($R_G \geq50$\,kpc), thus the evolution of mass and size is almost independent of concentration.

As shown in Fig.~\ref{ml-NFW} for the low halo concentration ($c= 5$) the half-mass radii of the star clusters evolves more similarly at the different galactocentric distances compared to the more concentrated haloes. The final 3D half-mass radii of all clusters orbiting in the range of 2--50\,kpc are distributed in a narrow range of 15--20\,pc. For models within the most concentrated of our haloes, the simulated clusters evolving in the inner part of the galaxy keep losing mass at a faster rate than the clusters orbiting in the less concentrated halo and hence their half-mass radius decreases towards the end of their lifetimes. Here in the inner part of the galaxy the effect of the DM halo concentration is most visible, that is, higher concentration means faster disruption and less expansion.

Figure~\ref{Rh-Rg} shows the 3D half-mass radius, $r_h$, of simulated star clusters orbiting at different distances from the galactic center, $R_G$, after 13 Gyr of evolution. In agreement with the functional form of hyperbolic tangent proposed by \citet{Madrid12},
\begin{equation}\label{eq:rh-RG}
r_h = r_{h,max}\cdot\tanh(\alpha R_G),
\end{equation}
where $r_{h,max}$ and $\alpha$ are two free parameters of the fit, we derived an $r_h-R_G$ relationship for each of our galaxy models. The best-fitting parameters we obtained are indicated in Table~\ref{fit-nfw}. It can be seen that the maximum size of all modeled clusters orbiting in the outer part of the galaxies are the same, meaning the value of $r_{h,max}$ is independent of halo concentration. \emph{However, the parameter $\alpha$ that defines the inner slope (within the inner 20 kpc) of this function is significantly different for various halo concentrations.} This is because the parameter $\alpha$ is the proxy of the tidal field. The concentrated mass of the galaxy inside a given galactocentric distance is different for different halo concentrations. This implies that the functional form of the $r_h-R_G$ relation (or, in other words, the onset of the plateau in Fig.~\ref{Rh-Rg}) for a given galaxy could potentially be interpreted as a tracer of the halo concentration of the host galaxy.

In the next Section we are going to test how a central baryonic component influences these findings.

\subsection{Evolution in a 3-component galactic model: two-parameter NFW profile} \label{ssec:NFW-2}

In this section we add a Milky Way-like stellar component (bulge and disk) to the NFW halo, producing a three-component model for the host galaxy. We use a mass of  $M_b=1.5 \times 10^{10}\msun$, and $M_d=5 \times 10^{10}\msun$ for the bulge and disk component, with $a=4$\,kpc (disk scale length) and $b=0.5$\,kpc (disk scale height) chosen to be similar to our Galaxy. We calculated models orbiting at different galactocentric distances, $R_G=$\,5, 8.5, 10, 15, 20, 30, 50, 100\,kpc.

The results of our simulations are illustrated in Fig.~\ref{ml-Rh-2}. As expected, mass and size of the clusters evolve similar to the models in Sec.~\ref{ssec:NFW}. As shown in the bottom panels of Fig~\ref{ml-Rh-2}, the initial increase in the half-mass radius, by a factor of 1.7, during the first 500\,Myr  is due to  dilution of the cluster's gravitational potential well, driven by the early impulsive mass-loss associated with stellar evolution \citep{Shin13}.  After this initial rapid evolution, the cluster evolution depends on the properties of the galactic potential. For the innermost clusters, stars are tidally-stripped away from the  cluster quickly after the initial expansion. This leads to a faster disruption and a smaller half-mass radius after 13\,Gyr of evolution. See, e.g., the cluster orbiting at $R_G=5$\,kpc, or at $R_G=15$\,kpc in the galaxy with $M_{vir} = 10^{13}\msun$ and $c=25$. In contrast,  the outermost models keep expanding, losing fewer stars to tidal stripping during the 13\,Gyr of evolution owing to the shallower potential of the galaxy at large galactocentric distances.

The tidal field in the inner part of the 3-component Milky Way-like galaxy \textbf{($R_G\leq 10$ kpc)} is governed by the stellar components rather than by the dark matter halo. That is, the presence of the stellar components plays a dominant role for clusters evolving in the inner part of the low-mass halo. In fact, the main factor is the combined, enclosed mass of the dark matter halo and the stellar component at a given orbital distance from the center of the galaxy. Therefore, the evolution of star clusters orbiting at small $R_G$ (see, e.g., the models evolving at $R_G=5$ kpc in Fig.~\ref{ml-Rh-2}) is not very sensitive to the halo parameters, because the ratio of stellar mass to dark matter mass is larger at small galactocentric distances.

The final values of the half-mass radii of our cluster models after a Hubble time of evolution at a given galactocentric distance are shown in Fig.~\ref{Rh-Rg-2}. Similar to the results of Sec.~\ref{ssec:NFW} we derived a $r_h-R_G$ relationship for the modeled clusters in the mathematical form of a hyperbolic tangent (Eq.~\ref{eq:rh-RG}). The two best-fitting parameters we obtain for all calculated models are given in Table~\ref{fit-nfw}. \emph{For all models  $r_{h,max}\approx20$\,pc is obtained for the maximum half-mass radius, independent of concentration or viral mass of the halo.} However, the onset of the plateau or, correspondingly, the slope within the inner 10-20\,kpc, which is defined by the parameter $\alpha$, correlates with the structural parameters of the halo. This is because clusters orbiting at distances beyond $R_G\approx 50$\,kpc evolve as if they are in isolation. Therefore, the mass-loss and size expansion is driven by the two-body relaxation process alone.

In contrast, the inner slope of this function differs for different virial masses and concentration parameters of the halo. Especially, this difference is evident for the more dark-matter dominated galaxy (i.e., $M_{vir} = 10^{13}\msun$). As illustrated in the right panel of Fig.~\ref{Rh-Rg-2}, the more concentrated halo results in a shallower slope $\alpha$ within the inner 20\,kpc in the $r_h-R_G$ relation. This means that the radius of the cluster increases more slowly with increasing galactocentric distance, or in other words, the onset of the plateau in the $r_h-R_G$ relation occurs at a larger galactocentric distance. This difference comes from the fact that the enclosed mass of the DM halo at a given galactocentric distance in the more concentrated halo model is larger than that of less concentrated ones. Thus the galactic tidal field is stronger in the more concentrated galaxy leading to the enhanced mass-loss driven by the galactic tide, and the stronger cut-off it inflicts on the clusters.

As mentioned in Sec.~\ref{sec:potential}, cosmological simulations of structure formation find that DM halo concentrations and virial masses are correlated. In the next Section we are going to test how such a correlation affects our ability to infer information on the DM halo of a galaxy when its DM content is unknown.

\begin{figure*}
\centering
\includegraphics[width=85mm]{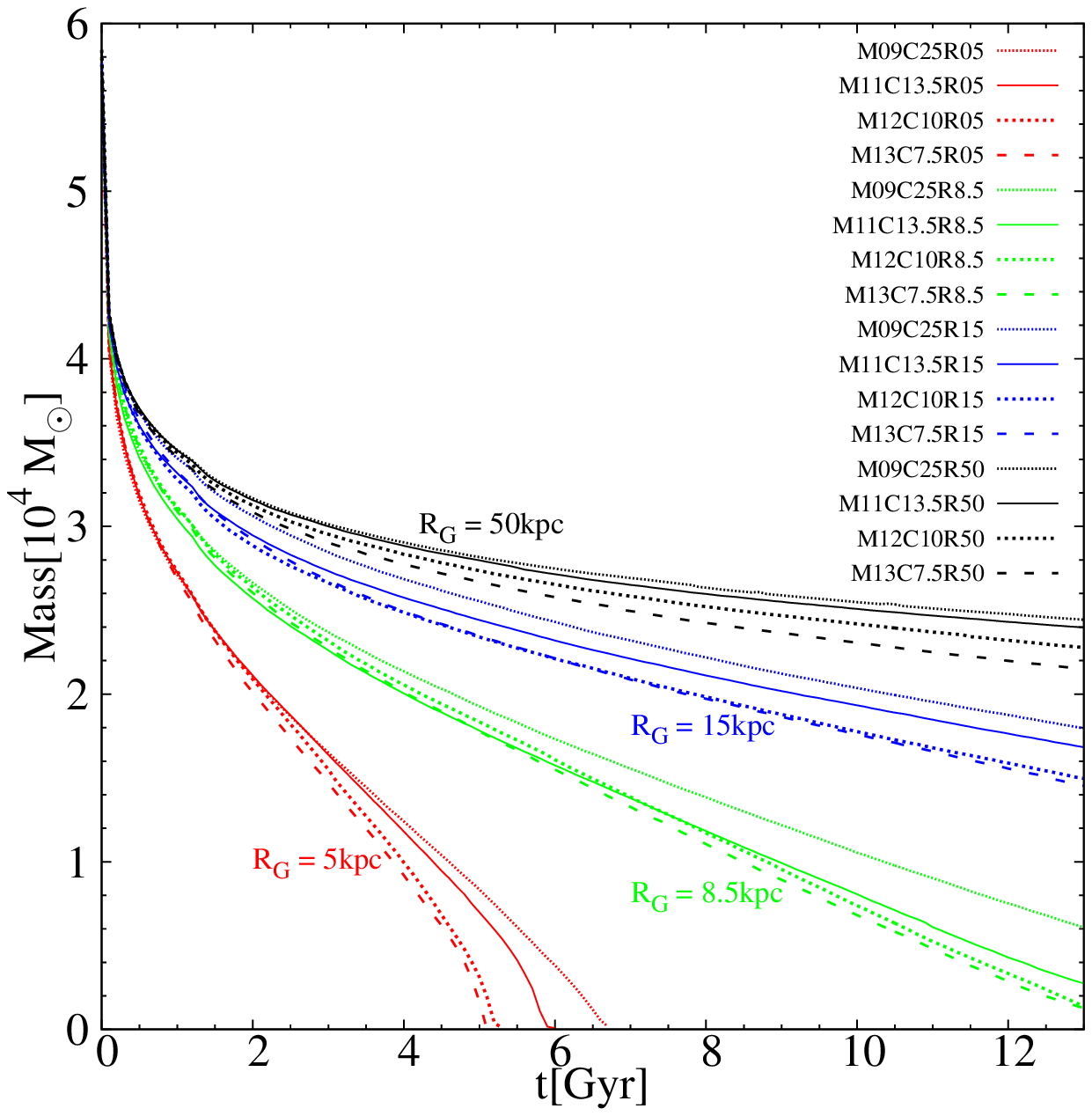}
\includegraphics[width=85mm]{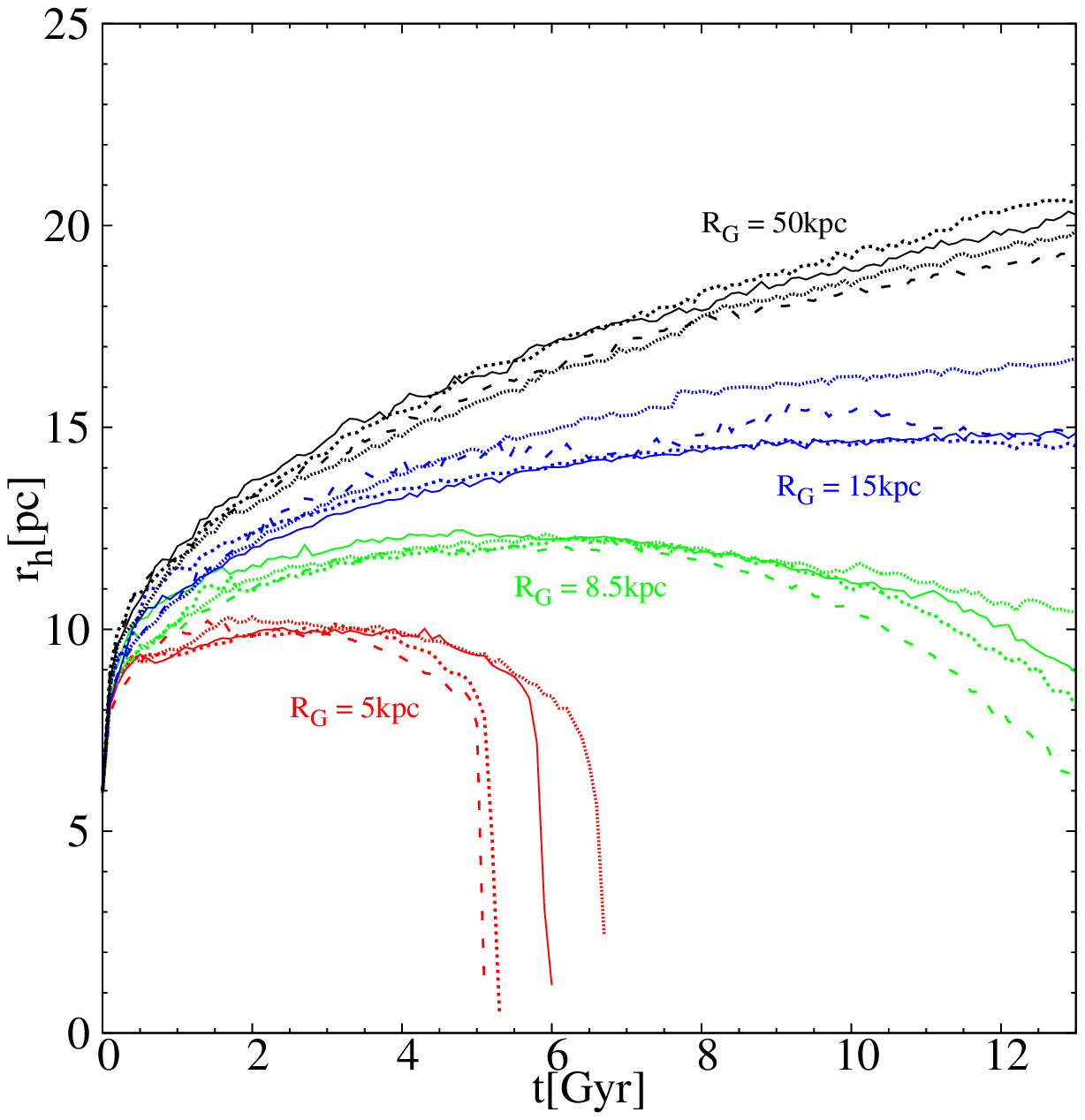}
\caption{Mass-loss and half-mass radius evolution with time as in Fig.~\ref{ml-Rh-2}, but here models are orbiting in a one-parameter halo model, where concentration and virial mass are related following Eq.~\ref{eq:c-mvir-relation}. All star clusters start with a 3D half-mass radius of 6\,pc and $N=10^5$ stars.  Different orbital distances are indicated in the panels.  All models orbiting at galactocentric distances $\leq8.5$\,kpc have lost all of their mass within a Hubble time. At a given galactocentric distance, we see that there exists little difference between the half mass radius and mass-loss rate of clusters that evolve in galaxies with different halo masses.} \label{ml-1}
\end{figure*}

\begin{figure}
\centering
\includegraphics[width=90mm]{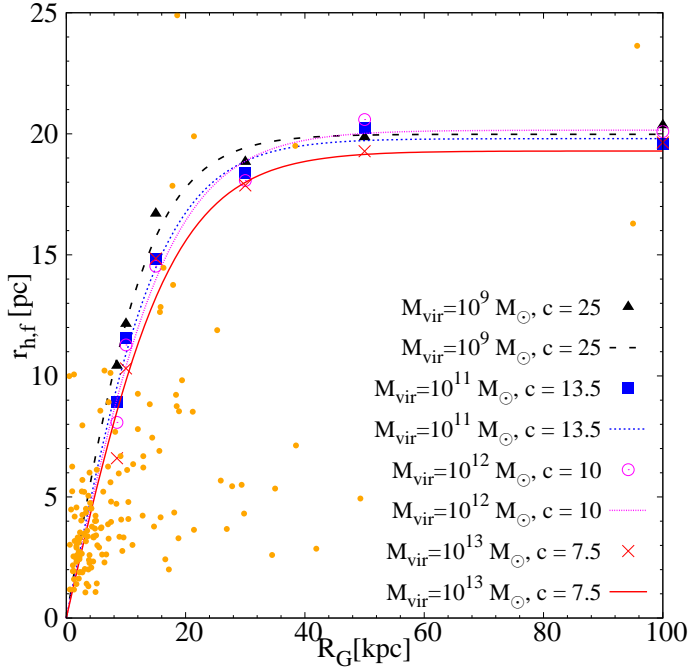}
\caption{Final 3D half-mass radii of simulated star clusters at different galactocentric distances in the one-parameter halo models with different viral masses and concentrations following the $M_{vir}-c$ relation (Eq.~\ref{eq:c-mvir-relation}). The lines are the best-fitting functions given by Eq.~\ref{eq:rh-RG}. The $r_h-R_G$ relations are similar for all halo models despite their very different virial masses. As in Fig.~\ref{Rh-Rg-2}, the 3D half-mass radii and Galactocentric distances of the MW GCs are over-plotted as filled circles for comparison.}
\label{Rh-Rg-1}
\end{figure}

\begin{table*}
\centering
\caption{The initial and final parameters of the $N$-body runs and of the dark matter halo as in Table~\ref{tab-nfw}, but for the one-component NFW models (Sec.~\ref{ssec:oneparameter}).}
\begin{tabular}{ccccccccccccc}
\hline
(1)&(2)&(3)&(4)&(5)&(6)&(7)&(8)&(9)&(10)\\
\hline
$R_{G}$ &        $M_{tot}$      & $\bar{\rho}_G$      &   $f_0$       & $M_{vir}$   &$c$&$r_{vir}$&$r_{s}$&$r_{h,f}$&  $M_{f}$ \\
\,[kpc] & [$10^{10}M_\odot]$ &$[M_\odot pc ^{-3}]$ & $[r_h/r_t]_0$ & [M$_\odot$] &         & [kpc]   & [kpc] & [pc]    & [$10^3$M$_\odot$] \\
\hline
5     & 3.6  & $6.9 \times 10^{-2}$ & $0.134$  &$10^{9}$  & $25$ & $21$ & $0.8$ &--      &--      \\
8.5   & 5.0  & $1.9 \times 10^{-2}$ & $0.092$  & $10^{9}$ & $25$ & $21$ & $0.8$ & $10.4$ & $6$ \\
10    & 5.4  & $1.3 \times 10^{-2}$ & $0.081$  & $10^{9}$ & $25$ & $21$ & $0.8$ & $12.2$ & $10$  \\
15    & 6.0  & $4.2 \times 10^{-3}$ & $0.057$  & $10^{9}$ & $25$ & $21$ & $0.8$ & $16.7$ & $18$  \\
30    & 6.5  & $5.7 \times 10^{-4}$ & $0.029$  & $10^{9}$ & $25$ & $21$ & $0.8$ & $18.8$ & $23$  \\
50    & 6.6  & $1.3 \times 10^{-4}$ & $0.018$  & $10^{9}$ & $25$ & $21$ & $0.8$ & $19.9$ & $24$  \\
100   & 6.7  & $1.6 \times 10^{-5}$ & $0.009$  & $10^{9}$ & $25$ & $21$ & $0.8$ &  $20.3$& $25$  \\
\hline
5     & 4.3   & $8.2 \times 10^{-2}$ & 0.139     &$10^{11}$ & 13.5 & 93 & 6.9  &--   & -- \\
8.5   & 6.4   & $2.5 \times 10^{-2}$ & 0.097     & $10^{11}$& 13.5 & 93 & 6.9  & 8.9 & 2 \\
10    &7.0    & $1.7 \times 10^{-2}$ & 0.086     &$10^{11}$ & 13.5 & 93 & 6.9  &11.6 & 8\\
15    & 8.5   & $6.1 \times 10^{-3}$ & 0.062     &$10^{11}$ & 13.5 & 93 & 6.9  &14.8 & 16 \\
30    & 11.2  & $1.0 \times 10^{-3}$ & 0.034     &$10^{11}$ & 13.5 & 93 & 6.9  &18.4 & 22 \\
50    & 13.5  & $2.6 \times 10^{-4}$ &0.022      &$10^{11}$ & 13.5 & 93 & 6.9  &20.3 & 23\\
100   & 16.8  & $4.0 \times 10^{-5}$ & 0.011     &$10^{11}$ & 13.5 & 93 & 6.9  &19.6 & 24 \\
\hline
5     & 5.1    & $9.7 \times 10^{-2}$    & 0.143 &$10^{12}$ & 10    &200 & 20.1 &--& --   \\
8.5   &8.7     & $3.4 \times 10^{-2}$    & 0.102 &$10^{12}$ & 10    &200 & 20.1 & 8.1 & 1\\
10    &  10.1  & $2.4 \times 10^{-2}$    & 0.092 &$10^{12}$ & 10    &200 & 20.1 &11.3 & 7  \\
15    & 14.6   & $1.0 \times 10^{-2}$    & 0.069 &$10^{12}$ & 10    &200 & 20.1 &14.5 &  14  \\
30    &  27.4  & $2.4 \times 10^{-3}$    & 0.043 &$10^{12}$ & 10    &200 & 20.1 &18.1 & 21 \\
50    & 42.4   & $8.1 \times 10^{-4}$    &0.030  &$10^{12}$ & 10    &200 & 20.1 &20.6 & 22 \\
100   & 70.6   & $1.7 \times 10^{-4}$    & 0.018 &$10^{12}$ & 10    &200 & 20.1 &20.1 & 24 \\
\hline
5     & 6.2    & $1.2 \times 10^{-1}$    & 0.149  &$10^{13}$ &7.5   & 439 & 57.7  &--   & -- \\
8.5   & 12.1   & $4.7 \times 10^{-2}$    & 0.109  &$10^{13}$ &7.5   & 439 & 57.7  & 6.6 & 1 \\
10    & 14.9   & $3.6 \times 10^{-2}$    & 0.098  &$10^{13}$ &7.5   & 439 & 57.7  &10.3 & 6 \\
15    & 25.6   & $1.8 \times 10^{-2}$    & 0.077  &$10^{13}$ &7.5   & 439 & 57.7  &14.9 &  14 \\
30    & 67.3   & $6.0 \times 10^{-3}$    & 0.053  &$10^{13}$ &7.5   & 439 & 57.7  &17.9 & 20\\
50    & 133.0  & $2.6 \times 10^{-3}$    & 0.040  &$10^{13}$ &7.5   & 439 & 57.7  &18.6 & 21\\
100   & 301.0  & $7.2 \times 10^{-4}$    & 0.028  &$10^{13}$ &7.5   & 439 & 57.7  &19.6 & 23 \\
\hline
\end{tabular}
\label{tab-1}
\end{table*}

\subsection{Evolution in a 3-component galactic model: one-parameter NFW profile}\label{ssec:oneparameter}

As mentioned in Sec.~\ref{sec:potential}, cosmological simulations have shown that virial mass, $M_{vir}$, and concentration, $c$, of NFW-like DM haloes, are correlated as in Eq.~\ref{eq:c-mvir-relation}, such that the concentration decreases as the virial mass increases. This leaves us, in principle, only one parameter, $M_{vir}$, to characterize the haloes of our model galaxies.

In this section we describe the results from simulations of star clusters evolving within a 3-component galaxy model including such a one-parameter NFW halo. We use a grid of four different virial masses:  $M_{vir}=10^9, 10^{11}, 10^{12}$, and $10^{13}\msun$. According to Eq.~\ref{eq:c-mvir-relation}, the corresponding concentrations are $c=25$, 13.5, 10, and 7.5, respectively.

As in Sec.~\ref{ssec:NFW-2}, the same simulations were carried out at different galactocentric distances for each halo model. To show how changes in the galactic halo potential can affect the cluster's evolution, Fig.~\ref{ml-1} displays the total mass of stars bound to the cluster and the evolution of the half-mass radius over time for all 21 simulations. The left panel of Fig.~\ref{ml-1} shows that, again, the early expansion is the same in all models due to the same internal initial conditions. But, since the long-term cluster mass loss for these clusters can be regarded as a runaway overflow over the tidal boundary, the models evolving in the inner part of the galaxies disrupt faster than the outer-halo models. Due to the large galactocentric distances, the outermost clusters are initially extremely under-filling their tidal sphere and, hence, can expand more than the models in the innermost galaxy.

It can be seen that the final sizes of models evolving in different DM haloes with a wide range of virial masses (i.e., from $M_{vir}=10^9$ to $10^{13}\msun$) are very similar to each other. This interesting similarity is also illustrated in Fig.~\ref{Rh-Rg-1}, where we show the final 3D half-mass radius versus galactocentric distance of the simulated star clusters.  As given in Table~\ref{fit-nfw}, the fits to the $r_h-R_G$ relations are nearly identical for all haloes with different virial masses. This trend is due to the fact that the models with higher virial mass have a lower concentration, and hence at a given $R_G$ the enclosed halo mass is almost the same for all models. The small differences in the $r_h-R_G$ relations indicate that \emph{the shape of the globular-cluster size distribution should be nearly independent of the virial mass of a galaxy's DM halo, if the $M_{vir}-c$ correlation for DM haloes from cosmological simulations is valid.}

\section{Discussion}\label{sec:discussion}

The findings of our three sets of models can be understood by keeping in mind that all our clusters start with the same initial conditions, and by looking at the similarities and differences of the galactic density profiles. In the following we are going to look at our data using filling factors, which are dimensionless measures of the importance of tidal effects for the clusters.

\subsection{Filling factors}

\begin{figure*}
\centering
\includegraphics[width=170mm]{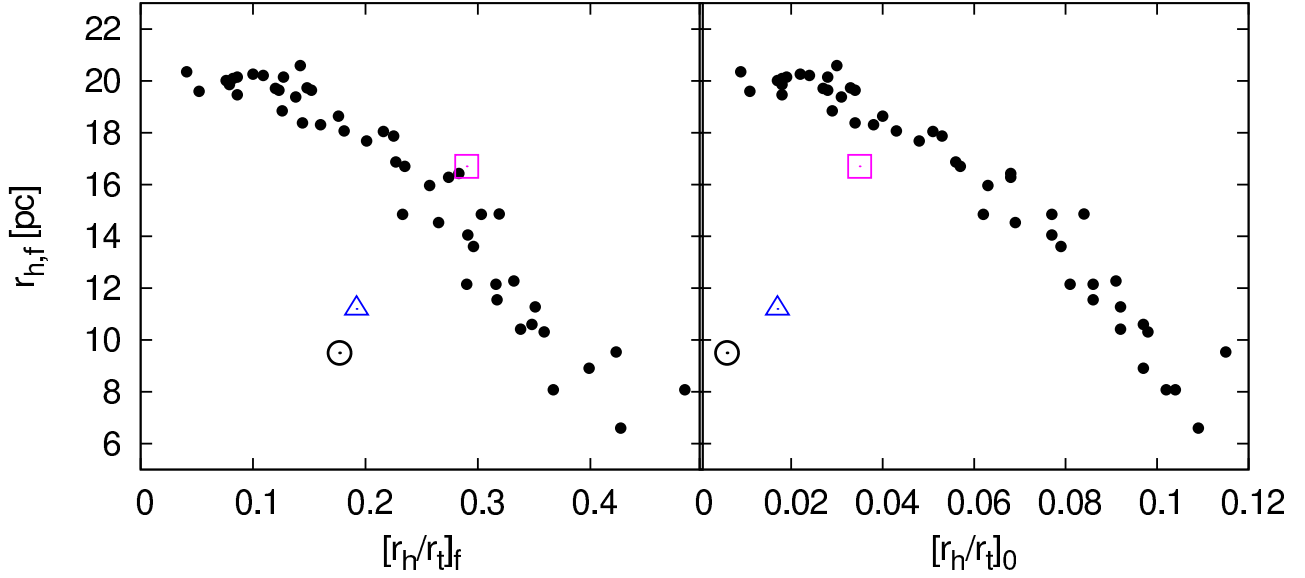}
\caption{Final 3D half-mass radius ($r_{h,f}$) of the simulated star clusters after a Hubble time of evolution for all surviving models versus the initial (left panel) and final (right panel) filling factors plotted for different galactic potentials. Especially apparent is the fact that for a given initial half-mass radius ($r_{h}=6$\,pc) and number of stars  ($10^{5}$), the final size of the star clusters is nearly determined by their initial/final filling factor. Three additional models orbiting at galactocentric distance of $R_{G} = 26$\,kpc with a lower number of initial particles of $N=41000$ and initial 3D half-mass radius of 0.5, 1.5, and 3\,pc are represented with an open circle, a triangle and a square, respectively.}
\label{Rh-filling}
\end{figure*}

In Fig.~\ref{Rh-filling} we show the final sizes of our modeled star clusters versus their initial and present-day ($f_0$ and $f_f$) filling factors $(f = r_h/r_t)$. As can be seen, the cluster final sizes decrease for initially tidally-filling models (i.e., $f_0 \geq 0.04$). That is, sizes decrease for models spilling over the tidal boundary at some point during their expansion. Therefore, their final half-mass radii are limited by the tidal field. Cluster with initial filling factors of about 0.1 and larger do not expand at all over a Hubble time as any expansion results in direct mass loss for these models. On the contrary, cluster that are strongly tidally-underfilling at the beginning of the simulation (i.e., $f_0 \leq 0.04$) have all more or less the same final half-mass radius after a Hubble time. This is due to the fact that, at the end of the simulations, they are still not filling their Roche lobes, i.e., they are still in the expansion phase.

The same trend is obtained for the present-day filling factors. The relation for $f_f$ is similar to the one for $f_0$ but shifted to higher values of $r_h/r_t$. This similarity is due to the fact that the initial conditions of all our models are the same and only the tidal field is varied from model to model. Thus, our findings will be valid for GCs that were born with similar configurations. However, for a comprehensive understanding of cluster sizes in galaxies, we will have to investigate wide distributions of initial conditions.

\citet{Haghi14} showed how sensitive the size evolution of star clusters is to the assumed initial conditions of the clusters. Here, we carried out three further simulations with a lower initial number of particles ($N=41000$), and thus lower initial mass of $M = 22000\msun$. The models started with different initial half-mass radii of 0.5, 1.5, and 3\,pc. They orbit at a galactocentric distance of $R_{G} = 26$\,kpc in a MW-like galactic potential with an NFW halo. After 13\,Gyr of evolution they reached final 3D half-mass radii of $r_{h,f}=9.5$, 11.2 and 16.7\,pc, respectively. The final sizes of these three clusters are indicated by open symbols in Fig.~\ref{Rh-filling}. According to the initial tidal radius of $r_t=84$ pc, the initial filling factors of these models are 0.006, 0.017, and 0.035. As can be seen, the final half-mass radii of these models do not follow the trend between half-mass radius and filling factor that we obtained for our original set of clusters, and somehow reflect the initial conditions of these clusters. Thus, each set of initial conditions will have its own $r_h-R_G$ relation. For the interpretation of observations of GC sizes it is therefore important to either take the spread in birth conditions of GCs into account, or to compare GCs that were born with similar properties.

In the following we will show that we can use the tight relation between filling factors and final half-mass radii to relate the cluster sizes to the mean enclosed mass density of the host galaxy.

\begin{figure*}
\centering
\includegraphics[width=170mm]{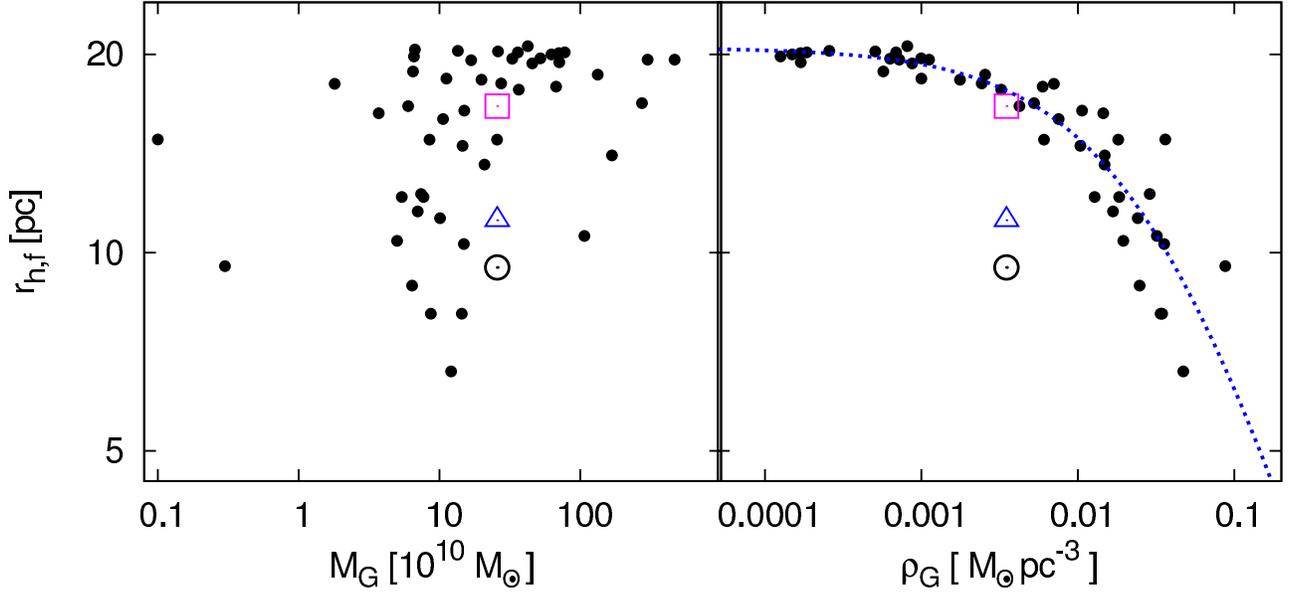}
\caption{Final 3D half-mass radius ($r_{h,f}$) of the simulated star clusters after a Hubble time of evolution for all  surviving models versus the enclosed galactic mass ($M_{tot}(R_G)$) (left panel), and versus the enclosed mean mass density of the host galaxy (right panel), defined as $\rho_{G}=3M_{tot}(R_G)/(4\pi R_G^3)$, plotted for different galactic halo potentials. The final sizes of the star clusters are closely related to the enclosed mean mass density of the galaxy rather than the enclosed galactic mass. The blue dashed line is the best-fitting function (Eq.~\ref{fit}) to the modeled clusters. Three additional models at galactocentric distance of $R_{G} = 26$\,kpc with a lower number of initial particles of $N=41000$ and initial 3D half-mass radius of 0.5, 1.5, and 3\,pc are represented with an open circle, a triangle and a square, respectively.}
\label{Rh-density}
\end{figure*}

\begin{figure}
\centering
\includegraphics[width=87mm]{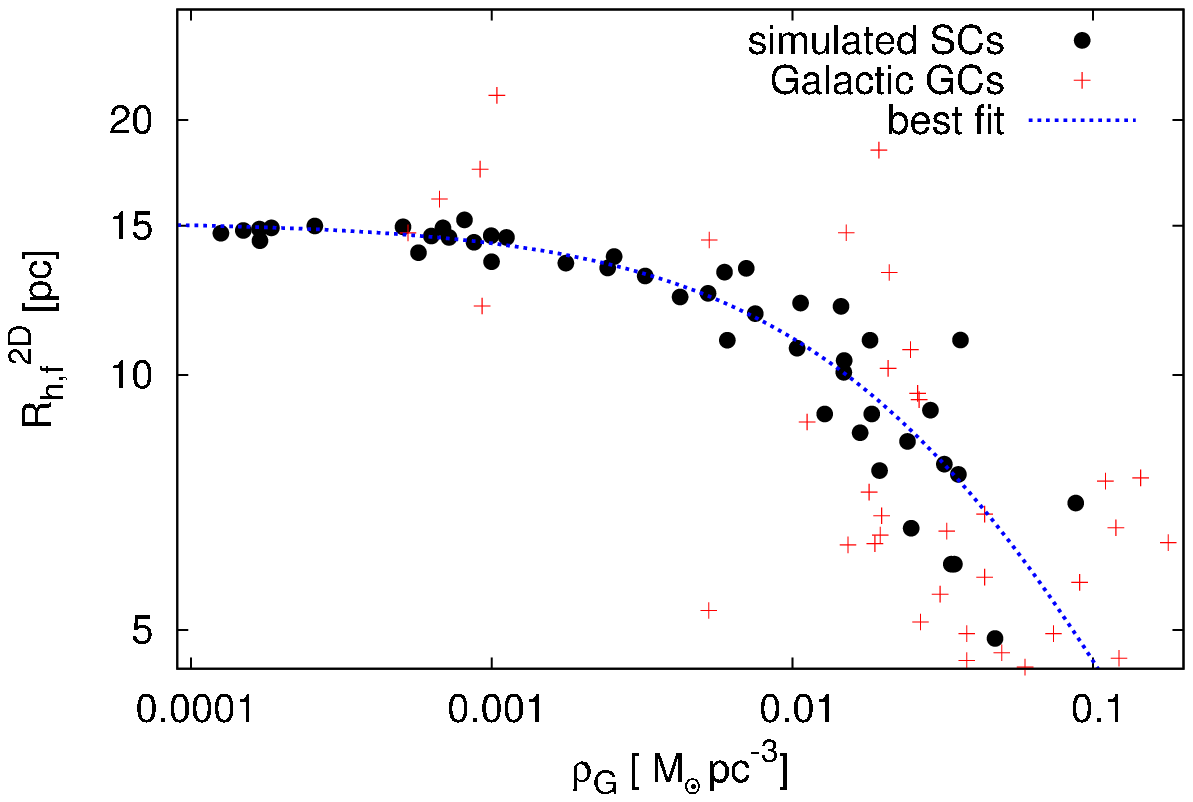}
\caption{Final 2D half-mass radius of simulated star clusters ($r_{h,f}$) after a Hubble time of evolution for all calculated models versus the enclosed mean mass density of the host galaxy. Data from \citet{Harris10} of the MW GC population in the range covered by our modeled star clusters (i.e., clusters with $R_{h,f}\geq4.5$ pc and orbiting at galactocentric radii $\geq5$\,kpc) are over-plotted. The blue dashed line is the best-fitting function given by Equation \ref{fit} to the modeled clusters.}
\label{MWdata}
\end{figure}

\subsection{Final sizes}

\begin{equation}
f^3\propto r_t^{-3} \propto M_{tot}(R_G)/R_G^3 \propto \bar{\rho}_G.
\end{equation}

Figure~\ref{Rh-density} shows the final sizes of our modeled star clusters versus the enclosed galaxy mass, $M_{tot}(R_G)$, calculated by integrating over the assumed halo density profile and adding the masses of the baryonic components if applicable. We also show the same data plotted against the mean mass density of the host galaxy inside the corresponding galactocentric distance of each cluster ($\rho_G = \frac{3M_{tot}(R_G)}{4 \pi R_G^3}$). There is no clear correlation between $r_{h,f}$ and $M_{tot}$. Instead, the final sizes of star clusters can be determined uniquely by the galactic mean mass density within the galactocentric distance of a given orbiting cluster.  The physical interpretation of this proportionality lies in the meaning of the
galactic density for the zero-energy surface (the tidal radius) of the cluster:
\begin{equation}
r_t = \left(\frac{GM_{cl}}{\Omega^2-\partial^2\phi/\partial R^2}\right)^{1/3},
\end{equation}
where $M_{cl}$ is the cluster mass, and $\Omega$ is its angular velocity in its
orbit around the galaxy \citep{King62}. Poisson's equation relates the second
derivative of the gravitational potential, $\phi$, to the Galactic density
\begin{equation}
\partial^2\phi/\partial R^2 = 4\pi G\rho_G.
\end{equation}
We can use this proportionality to relate the tidal radius to the galactic density.
In other words, using
\begin{equation}
r_t =R_G\left(\frac{M_{cl}}{2M_{tot}(R_G)}\right)^{1/3}
\end{equation}
we can show that the cubic of a cluster's filling factor is proportional to the mean
density of the galaxy enclosed within the cluster's galactocentric radius:

\begin{equation}
f^3\propto r_t^{-3} \propto M_{tot}(R_G)/R_G^3 \propto \bar{\rho}_G.
\end{equation}

We can therefore use the distribution of half-mass radii to infer the underlying galactic density profile. The least-squares fit to the data points that we show in the right panel of Fig.~\ref{Rh-density} is obtained using the functional form
\begin{equation}  \label{fit}
r_{h,f} = \frac {r_{h,max}}{1+(\frac{\rho_G}{\rho_c})^{n}},
\end{equation}
where $\rho_c$ and $n$ are arbitrary coefficients. The best-fitting  coefficients are found to be $r_{h,max}=20.50$ pc, $\rho_c=0.03 \msun$\,pc$^{-3}$, and $n=0.8$.

In Fig.~\ref{MWdata}, we compare the two-dimensional half-mass radii of our simulations with observational data of the MW GC population taken from \citet{Harris10}. In order to calculate the mean mass density within the present-day Galactocentric distances of the MW clusters, we use $M_{vir}=1.69\times 10^{12}\msun$ and $c=5.1$ from \citet{Kuepper15} for the parameters of the NFW halo. Since our modeled clusters in galaxies with a disk and a bulge are orbiting at galactocentric radii $\geq5$\,kpc, we exclude MW clusters with radii $\leq5$\,kpc  for our comparison. As can be seen, if only clusters above a size of $R_{h,f}=5$\,pc (which is about the range of our modeled star clusters) are taken into account, the actual observed data of MW GCs roughly follows the theoretical curve. Some outliers above the curve are probably tidally overflowing clusters like Palomar\,5 and Palomar\,14. Outliers below the curve are either still tidally under-filling or on eccentric orbits within the MW. With a better understanding of such outliers, our method could be used as a powerful tool to measure gravitational potentials of galaxies.

However, our initial question remains unanswered so far: can we use an ensemble of GC sizes and their galactocentric distances to infer the underlying gravitational potential of the host galaxy by fitting a $\tanh$-like function to the data?

\subsection{Fit parameters of the $r_h-R_G$ relation}

\begin{figure*}
\centering
\includegraphics[width=160mm]{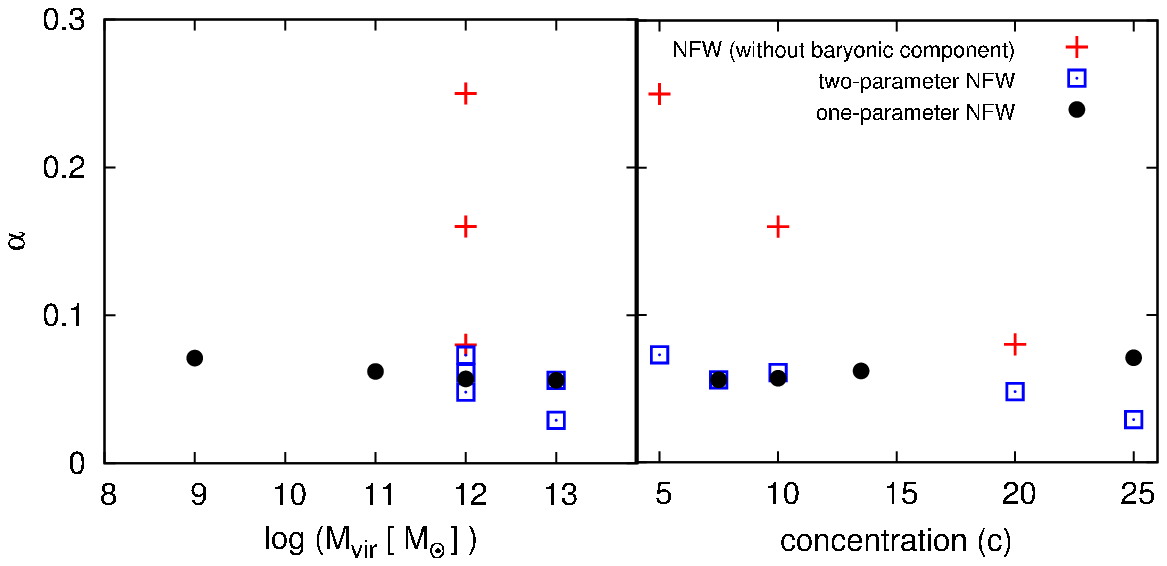}
\includegraphics[width=160mm]{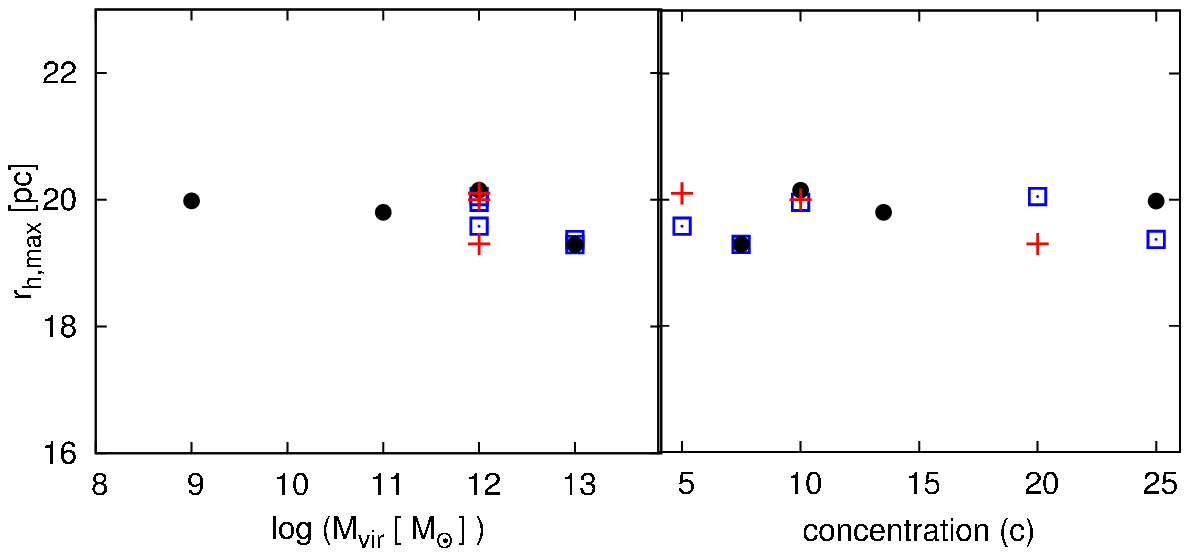}
\caption{Upper panels: slope of the $r_{h,f}-R_G$ relation ($\alpha$) compared to the NFW halo parameters ($M_{vir}$ and $c$). For the one-parameter NFW halo (black filled circles), where $M_{vir}$ and $c$  are related to each other as suggested by cosmological simulations, $\alpha$ is basically independent of the halo parameters. The same holds for the models with a central, dominant baryonic component, since $\alpha$ reflects the slope of the $r_{h,f}-R_G$ relation in the inner part of the galaxies, where the baryons dominate the galactic mass. In the dark-matter only galaxies (red plus signs) the slope is decreasing with increasing concentration. Bottom panels:  maximum 3D half-mass radius of simulated star clusters ($r_{h,max}$) after a Hubble time of evolution versus the NFW halo parameters $M_{vir}$ and $c$. The maximum sizes of all modeled clusters orbiting in the outer part of the galaxies are the same, meaning the value of $r_{h,max}$ is largely independent of the galaxy characteristics. It is set by the initial conditions of the GCs, which were chosen to be the same for all our models.}
\label{correlation}
\end{figure*}

Figure~\ref{correlation} depicts the fitting parameters of the $r_h-R_G$ relations given by Eq.~\ref{eq:rh-RG} ($\alpha$ and $r_{h,max}$) versus the NFW halo parameters ($M_{vir}$ and $c$) for our different galaxy models. In the upper panels of this figure we see that the slope of the $r_h-R_G$ relation, $\alpha$, shows a clear correlation with halo concentration in our dark-matter only galaxies. For these galaxy models, clusters orbiting in a low concentration halo have a steeper slope in the $r_h-R_G$ relation. This can be understood when considering that $\alpha$ reflects the increase in cluster sizes in the inner part of the galaxy. The concentration parameter of the halo determines the mean mass density within this inner part, and hence the sizes of the clusters. In the case of a one-parameter halo model there is no such sensitivity on the concentration parameter, since dark matter haloes following the $M_{vir}-c$ relation have similar mean densities within their centers. The (fixed) baryonic components (disk and bulge) help to keep this mean mass density constant when we vary the concentration and virial masses of the halo. The fit parameter $\alpha$ is therefore a sensitive tracer of the inner density profiles of galaxies, and galaxies with similar inner density profiles should therefore show similar $r_h-R_G$ relations.

According to the lower panels of Fig.~\ref{correlation}, the maximum sizes that clusters can reach after a Hubble time of evolution are nearly independent of the halo parameters. This is easy to understand: remote, i.e., tidally-underfilling, star clusters evolve like isolated clusters and two-body relaxation is the dominant mechanism in their long-term evolution, and hence in determining the sizes of the clusters. However, the final sizes of star clusters, and consequently the fitting parameter $r_{h,max}$ of the $r_h-R_G$ relation, depend on their birth configurations. The plateau of the $r_h-R_G$ relation therefore tells us something about the formation of globular clusters. To assess this valuable information, more comprehensive sets of models have to be
computed and, other, computationally less expensive methods like, for example, MOCCA \citep{Giersz13} or EMACCS \citep{Alexander12}, should be used for this
purpose. Studies like \citet{Leigh15} and \citet{Webb15} are important first steps towards a better understanding of the birth conditions of globular clusters. The authors showed that present-day properties like binary fractions and stellar mass functions can be used to constrain the formation properties of star clusters. Once the formation of GCs is better understood, the plateaus in half-mass radii distributions can be used to efficiently measure density profiles of galaxies, as we have demonstrated in this work.

\subsection{Complications}

We should note that all models in our data set were evolved on circular orbits. Clusters on eccentric orbits evolve like clusters on circular orbits at a galactocentric distance that reflects the average tidal field  (\citealt{Webb14b, Cai15}, K\"upper et al. 2015, in preparation). Therefore, one can expect the final half-mass radius of a cluster orbiting on an eccentric orbit  to be a good proxy for the mean density inside the time-varying galactocentric distance (i.e., between apogalactic and perigalactic distances).

Furthermore, we should point out that we have assumed a strongly simplified galactic potential, which is static and axis-symmetric. Since dark matter haloes assemble hierarchically over time, the size scale evolution of GCs should be considered in a triaxial, time-dependent galactic potential \citep{Renaud15, Haghi15a}.

\begin{table}	
\centering
\caption{Best-fit parameters of the $r_{h,f}-R_G$ relation (Eq.~\ref{Rh-Rg-2}) for different galaxy models.}
\begin{tabular}{cccccc}
\hline
NFW  halo &&&&&\\
\hline
$M_{vir}$ [M$_\odot$] & $10^{12}$&$10^{12}$&$10^{12}$ &&\\
$c$ & 5 & 10 & 20&& \\
$r_{h,max}$ [pc] & 20.1&20.0 & 19.3&& \\
$\alpha$ & 0.25 & 0.16 & 0.08&&\\
\hline
two-component halo &&&&&\\
\hline
$M_{vir}$ [M$_\odot$] & $10^{12}$&$10^{12}$&$10^{12}$&$10^{13}$&$10^{13}$\\
$c$ & 5 & 10 & 20 & 7.5 & 25 \\
$r_{h,max}$ [pc]& 19.6 & 20.0 & 20.0 & 19.3 & 19.4 \\
$\alpha$ & 0.07 & 0.06 & 0.05 & 0.06 & 0.03\\
\hline
one-component halo &&&&&\\
\hline
$M_{vir}$ [M$_\odot$] & $10^{9}$&$10^{11}$&$10^{12}$&$10^{13}$&\\
$c$ & 25 & 13.5 & 10 & 7.5 &\\
$r_{h,max}$ [pc] & 20.0 & 19.8 & 20.1 & 19.3& \\
$\alpha$ & 0.07 & 0.06 & 0.06 & 0.06&\\
\hline
\end{tabular}
\label{fit-nfw}
\end{table}

\section{Conclusions}\label{sec:conclusions}

We have used direct $N$-body simulations to study the dynamical evolution of star clusters orbiting within the tidal field of different galaxy models. In particular, we investigated how a cluster's half-mass radius and mass-loss rate evolve over its lifetime, and how this evolution changes when we change the properties of the galaxy's dark matter halo. We created a grid of more than 60 GC models at various galactocentric distances within haloes of different virial masses and concentrations. Our main conclusions are the following:
\begin{enumerate}
\item First, a set of simulations without stellar component was created in order to investigate the main effect of halo structural parameters on the evolution of star clusters. By running models within the halo with the same virial mass ($M_{vir}=10^{12}\msun$) but different concentrations ($c=5$, 10, and 20) we determined that, as the halo concentration increases the star cluster's mass-loss also increases.  The impact of the halo concentration on the GCs' mass-loss rate and size scale is significantly more evident in the inner regions of the host galaxy. The evolution and final size of star clusters at $R_G=20$\,kpc and beyond are nearly independent of the value of $c$. Therefore, the shape of the whole distribution of sizes (i.e., $r_{h,f}(RG)$) has information about the host galaxy potential that can be assessed by fitting a $\tanh$-like function to it (e.g., Eq.~\ref{eq:rh-RG}).

\item We next investigated a three-component model of the galactic tidal field, consisting of a Milky Way-like stellar bulge and disk and an NFW halo. Comparing the results of two different halo models with the masses of $M_{vir}=10^{12}\msun$ and $10^{13}\msun$, we showed that the effect of halo concentration on the evolution of star clusters and especially on the $r_h-R_G$ relation is more evident in more strongly dark-matter dominated galaxies, since the baryonic components dominate the inner parts of galaxies with less massive DM haloes.
\item Finally, we examined a more realistic, cosmologically motivated, galactic model in which the NFW halo parameters $M_{vir}$ and $c$ are directly correlated, as suggested by simulations of structure formation in a $\Lambda$CDM Universe, leaving the DM halo as a one-parameter profile. Our models show that different masses of the host galaxy halo have no significant effect on the evolution of star clusters, since a higher virial mass is compensated by a lower concentration. This is apparently in contrast with the expectation that the dissolution rate of star clusters are higher in more massive galaxies \citep{Madrid14, Miholics14}.  In contrast, we here show that clusters with initially the same size and mass reach nearly the same size distributions after a Hubble time of evolution independent of DM halo masses.
\end{enumerate}

We found that the determining factor in the final size of the clusters is the mean mass density of a galaxy within the cluster orbiting radius rather than its total mass. Globular cluster sizes are therefore powerful tracers of galactic density profiles. This implies that, even if $M_{vir}$ is large,
clusters could evolve as if they were in isolation. Vice versa, if $M_{vir}$ is small but the DM halos is concentrated, its globular clusters may be fully eroded within a Hubble time. This is in agreement with the recent observations that show massive ellipticals (such as NGC\,4889) have rich, radially extended GC systems, while some compact dwarf ellipticals such as M\,32 show a lack of GCs.  Our results confirm the recent conclusions by \cite{Brockamp14}, who have found by means of comprehensive $N$-body experiments that the fraction of eroded GCs is nearly 100 percent in very compact M\,32-like dwarf galaxies, while the rate of erosion is lower in the most massive and extended galaxies like NGC\,4889 or M\,87 with $>10000$ globular clusters \citep{Harris13, Mieske14, Harris15}.

We suggest that observed $r_h-R_G$ distributions of GCs in galaxies with different masses, ranging from compact dwarfs to giant ellipticals, can potentially be used to test the $M_{vir}-c$ relation for dark matter haloes. This relation is an important prediction from the $\Lambda$CDM cosmological model, and should hold for the parts of galaxies that are not baryon dominated. Globular clusters are abundant tracers of halo density profiles out to large radii, and can yield estimates even for low-mass galaxies. This method will be a valuable addition to existing methods for measuring halo concentrations such as gravitational lensing (e.g., \citealt{Postman12}).

\section*{Acknowledgements}
The authors would like to thank Nathan Leigh for fruitful and stimulating discussions of the manuscript. AHWK acknowledges support through Hubble Fellowship grant HST-HF-51323.01-A awarded by the Space Telescope Science Institute, which is operated by the Association of Universities for Research in Astronomy, Inc., for NASA, under contract NAS 5-26555. This work was made possible by the facilities of Graphics Processing Units at the Institute for Advanced Studies in Basic Sciences (IASBS).

\label{lastpage}

\end{document}